\documentclass[fleqn,toc]{PoS}

\usepackage[utf8]{inputenc}
\usepackage{amsmath}
\usepackage{bm}
\usepackage{mathtools}
\usepackage{enumitem}

\usepackage{tikz}
\usepackage{pgfplots}
\usetikzlibrary{positioning,arrows,patterns}
\usetikzlibrary{decorations.markings}
\usetikzlibrary{calc}

\newcommand{\tr}{\mathrm{tr}\,}

\renewcommand{\d}{\mathrm{d}}
\renewcommand{\vec}[1]{\mathbf{#1}}
\renewcommand{\Re}{\mathrm{Re}\,}
\newcommand{\fsl}[1]{\not{#1}}

\title{QED Corrections to Hadronic Observables}

\ShortTitle{QED Corrections to Hadronic Observables}

\author{\speaker{Agostino Patella}\\%
      Theoretical Physics Department, CERN, Geneva, Switzerland\\
      Centre for Mathematical Sciences, Plymouth University, Plymouth, PL4 8AA, UK\\
      E-mail: \email{agostino.patella@cern.ch}}

\abstract{
When aiming at the percent precision in hadronic quantities calculated by means of lattice simulations, isospin breaking effects become relevant. These are of two kinds: up/down mass splitting and electromagnetic corrections. In order to account properly for the latter, a consistent formulation of electrically-charged states in finite volume is needed. In fact on a periodic torus Gauss law and large gauge transformations forbid the propagation of electrically-charged states. In this talk I will review methods that have been used or proposed so far in order to circumvent this problem, while highlighting practical as well as conceptual pros and cons. I will also review and discuss various methods to calculate electromagnetic corrections to hadron masses and decay rates in numerical simulations.\\
\vspace{3mm}\\
CERN-TH-2017-019
}

\FullConference{34th annual International Symposium on Lattice Field Theory\\
		 24-30 July 2016\\
		 University of Southampton, UK}

\begin{document}

\section{Introduction}
\label{sec:intro}

Isospin is an approximate symmetry violated at the percent level by two effects:
up and down-quark mass difference and electromagnetic interactions. There are
three main reasons to include isospin-breaking effects in lattice simulations:
\begin{itemize}

   \item \textbf{Calculate pure isospin-breaking effect}, like \textit{hadron-mass
   splitting} or \textit{up and down-quark renormalized mass difference}.
   
   The lattice community has been active on the front of hadron-mass
   splitting since the exploratory studies in the electroquenched setup
   in~\cite{Duncan:1996xy,Basak:2013iw,Blum:2010ym,Portelli:2012pn}. The 2015
   BMW work~\cite{Borsanyi:2014jba} on baryon-mass splitting remains a landmark
   for the complexity of the simulations and for the rigorous treatment of
   finite-volume effects in presence of dynamical photons. The
   QCDSF-CSSM-RIKEN-Kobe (in the following, QCDSF for short) collaboration has
   reported in this conference on their hadron-mass splitting
   calculation~\cite{Horsley:2015vla,Horsley:2015eaa,Young:LAT2016}. The BMW and
   QCDSF work has been reviewed Liu's plenary talk in this conference and
   related proceedings~\cite{Liu:2016kbb}.
   
   The BMW collaboration has also reported in this conference on their
   calculation of the up and down-quark mass difference, in the
   electroquenched setup~\cite{Fodor:2016bgu,Varnhorst:LAT2016}. The $m_u=0$
   solution to the strong CP problem is excluded by their calculation, assuming
   that the estimate for the systematic error due to electroquenching be
   reliable.

   All calculations mentioned so far of the BMW and QCDSF collaborations rely on
   non-local prescriptions to describe charged particles. I will come back to
   this point later on.

   \item \textbf{Improve on the determination of observable, which have reached
   the precision of order $\mathbf{1\%}$.} This is the case for \textit{decay rates of
   light mesons}.
   
   In fact the FLAG working group reports the following world averages for pion
   and kaon decay constants at the isosymmetric point~\cite{Aoki:2016frl}
   \begin{gather} F_{\pi} = 130.2(1.4) \text{ MeV} \nonumber \ , \\ F_{K} =
   155.6(0.4) \text{ MeV} \nonumber \ , \end{gather} with a relative error of
   $1\%$ and $0.3\%$ respectively. On the other hand radiative corrections to
   leptonic decay rates of pion and kaon are estimated to be about $1.8\%$ and
   $1.1\%$ respectively on the basis of $\chi$PT
   \cite{Rosner:2015wva,Cirigliano:2011ny}, and already dominate the error
   budget. In order to improve on the state-of-the-art lattice determinations of
   such decay rates, isospin breaking corrections must be included.
   
   The RM-SOTON collaboration has presented in this conference a strategy to
   calculate leptonic decay rates, and some preliminary results for light
   mesons in the electroquenched
   setup~\cite{Carrasco:2015xwa,Lubicz:2016xro,Lubicz:2016mpj,Tantalo:2016vxk}.
   Calculations of radiative correction to decay rates present extra
   complications with respect e.g. to mass-splitting calculations due to the
   existence of IR divergences in QED. Even though IR divergences have been
   studied since the early days of QED and are very well understood, they do not
   belong to the standard toolbox of the lattice community. For this reason I
   have chosen to devote section~\ref{sec:Smatrix} to reviewing general aspects
   of decay rates and IR divergences, in relation to the RM-SOTON calculation,
   but also with a (superficial) eye to heavy-meson decay rates.
   
   \item \textbf{Improve on theoretical estimates of possibly large radiative
   corrections}, like to \textit{decay rates of heavy mesons}, or to the
   \textit{HVP contribution to $g_\mu-2$}.
   
   The RBC-UKQCD collaboration has presented in this conference some preliminary
   studies on the isospin-breaking correction to the HVP contribution to
   $g_\mu-2$~\cite{Boyle:2016lbc}.
   
   On the other hand no activity has been reported on equally-interesting
   radiative corrections to decay rates of heavy mesons. For instance radiative
   corrections to $B \to D \ell \nu$ are relevant for a precise determination of
   $|V_{cb}|$ at the upcoming Belle II experiment (see for instance
   \href{https://confluence.desy.de/download/attachments/34042032/belle2-note-0021.pdf?api=v2}{\texttt{BELLE2-NOTE-PH-2015-002}}),
   and are expected to be of order $3\%$ as reported in the 2016 PDG
   review~\cite{Olive:2016xmw}. A search in the references therein
   (e.g.~\cite{Bailey:2014tva,Aubert:2007qs,Adam:2002uw,Aubert:2008yv,Amhis:2012bh})
   reveals that radiative corrections are estimated very crudely by separating
   three contributions:
   \begin{enumerate}
   
      \item Short-distance contributions, due to photons coupling to the $W$.
      These contributions can be systematically accounted for by means of the
      OPE.
      
      \item Long-distance soft-photon contributions
      (\textit{inner-bremsstrahlung}) in loops and finale-state radiation. These
      are analytically calculable.
      
      \item Long-distance hard-photon contributions
      (\textit{structure-dependent} contributions). These are fully
      non-perturbative, and they are either neglected or estimated by saturating
      relevant matrix elements with a few resonances (see
      also~\cite{Bernlochner:2011bia}).
   
   \end{enumerate}
   
   Structure-dependent contributions may be enhanced by quasi-on-shell
   resonances and collinear quasi-divergences which arise when charged leptons
   are produced with large energy, and neglecting them may be not fully
   justified at the required level of precision (some of these issues will be
   highlighted in section~\ref{sec:Smatrix}). It is important to understand if a
   robust strategy can be designed to estimate these contributions from lattice
   simulations.

\end{itemize}

Moving to more numerical issues, two very general strategies have been proposed
in order to calculate isospin-breaking effects in lattice simulations, which
have been already reviewed in detail in plenary
talks~\cite{Tantalo:2013maa,Portelli:2015wna} of previous editions of the
Lattice conference:
\begin{itemize}

   \item \textit{Generate configurations in the isosymmetric limit and reweight
   them.} The reweighting factor can be included in a standard fashion as a
   ratios of fermionic determinants averaged over the photon field as originally
   proposed in~\cite{Ishikawa:2012ix,Aoki:2012st}. Alternatively the reweighting
   factor (along with the observables) can be expanded at first order in
   $m_d-m_u$ and $\alpha_\text{em}$ (\textit{RM123
   method}~\cite{deDivitiis:2011eh,deDivitiis:2013xla}). In practice one needs
   to insert $\bar{\psi} \psi$ and $\bar{\psi} \gamma_\mu \psi$ operators into
   the original observables, multiply times the photon propagator and sum over
   the photon momenta. The convolution with the photon propagator can be
   stochastically estimated. The clear advantage of this particular incarnation
   of reweighting is that one has direct access to the leading isospin-breaking
   corrections as $O(1)$ observables, rather than $O(\alpha_\text{em}) \sim
   O(\Delta m_{ud})$ effects on $O(1)$ observables.
   
   \item \textit{Generate non-isosymmetric configurations at unphysically large
   values of $m_d-m_u$ and $\alpha_\text{em}$ and extrapolate to the physical
   point.} This strategy has been used by the BWM~\cite{Borsanyi:2014jba} and
   the QCDSF collaboration~\cite{Horsley:2015vla} (first explorations in the
   electroquenched setup can be found in~\cite{Duncan:1996xy}). As observed by
   the BMW collaboration $O(\alpha_\text{em}^2)$ effects seem to be surprisingly
   small if $\alpha_\text{em}$ is the renormalized coupling, and this is
   essential for the strategy to work. The extrapolation can be performed along
   different parameter trajectories. For instance the QCDSF collaboration
   claims~\cite{Horsley:2015eaa} that choosing a trajectory that departs
   symmetrically (in some sense that is specified in their work) from the
   SU($3$) symmetric point is particularly beneficial.

\end{itemize}
A detailed comparison between these two methods would be very interesting and is
still missing. A first move in this direction has been presented by the
RBC-UKQCD collaboration~\cite{Boyle:2016lbc} in the context of the
electroquenched approximation.

The inclusion of electromagnetic corrections in lattice simulations is
especially challenging from the theoretical point of view. Typically lattice
simulations assume a finite box with periodic boundary conditions along the
spatial directions. However Gauss law forbids states with total electric charge
different from zero in a box with periodic boundary conditions. This is an
essential obstruction for numerical simulations which aim at calculating
properties of charged particles. Several prescriptions have been proposed in
order to circumvent this problem. A big fraction of this contribution
(section~\ref{sec:charged}) is devoted to a critical review of these
prescriptions, what we know about them, together with some new results. In
particular I will discuss the following aspects:
\begin{itemize}

   \item If $\tilde{A}_\mu(p)$ is the photon field in Fourier space, the
   QED$_\text{TL}$ prescription~\cite{Duncan:1996xy} is defined by imposing the
   constraint $\tilde{A}_\mu(0)=0$. As already noticed by
   BMW~\cite{Borsanyi:2014jba}, QED$_\text{TL}$ does not have a transfer matrix
   with a regular perturbative expansion. QED$_\text{TL}$ has been used in the
   BMW calculation of the up and down-quark mass
   difference~\cite{Fodor:2016bgu,Varnhorst:LAT2016}).
   
   \item The QED$_\text{SF}$ prescription~\cite{Gockeler:1989wj} is defined by
   restricting the value of $V^{-1} e \tilde{A}_\mu(0)$ in the interval $(-\pi
   L_\mu^{-1},\pi L_\mu^{-1})$. Charged-field two-point functions in
   QED$_\text{SF}$ do not have a local representation in time, therefore the
   existence of a transfer matrix in the charged sector is not guaranteed (and
   very unlikely). QED$_\text{SF}$ has been used in all QCDSF calculations.
   
   \item The QED$_\text{L}$ prescription~\cite{Hayakawa:2008an} is defined by
   imposing the constraint $\tilde{A}_\mu(p_0,\vec{0})=0$ for any $p_0$.
   Renormalization of higher-dimensional operators by local counterterms breaks
   down in the $\phi^4$ scalar theory with a constraint analogous to
   QED$_\text{L}$. This also implies that the operator product expansion as well
   as the Symanzik effective-theory description of the lattice theory close to
   the continuum limit break down because of non-local contributions.
   QED$_\text{L}$ has been used in the BMW baryon-mass splitting
   calculation~\cite{Borsanyi:2014jba} and in the RM123-SOTON decay rate
   exploratory calculation~\cite{Lubicz:2016mpj}.
   
   \item QED with a massive photon (QED$_\text{m}$) is a perfectly consistent
   QFT. In finite volume the theory has two IR regulators, the photon mass
   $m_\gamma$ and the box size $L$~\cite{Endres:2015gda}. The $L \to \infty$
   limit has to be carefully taken before the $m_\gamma \to 0$ limit, as the two
   limits do not commute. For instance, if $m_\gamma$ is too small, the
   spatial-momentum dependence in charged-hadron two-point functions is suppressed
   and the extraction of states with definite momentum is hindered.
   
   \item QED with C-parity boundary conditions (QED$_\text{C}$) in the spatial
   directions is also a perfectly consistent
   QFT~\cite{Polley:1990tf,Lucini:2015hfa}. The boundary conditions partially
   break flavor conservation in a local fashion. The flavour mixing induced by
   the boundary conditions does not affect the majority of the stable hadrons,
   is exponentially suppressed with the volume in two-point functions, and
   absent in the renormalization of composite operators. QED$_\text{C}$ also
   allows a completely gauge-invariant description of charged interpolating
   operators.
   
\end{itemize}
The prescription that defines QED$_\text{L}$ has been widely believed to be
harmless for a few years, on the basis that operators with dimension not larger
than 4 renormalize as in infinite volume at one loop. The first sign of sickness
was analyzed in~\cite{Fodor:2015pna}: in the non-relativistic limit charged
particles and antiparticles do not decouple. The breakdown of the Symanzik
expansion casts a shadow on the ability to extract a reliable continuum limit,
especially when large values of $\alpha_\text{em}$ are used in the
extrapolation.\footnote{The continuum limit with QED deserves a special
discussion. I am not concerned here with possible non-perturbative definitions
and UV-completions of QED. In infinite-volume QED, $n$-point functions and
transition probabilities have a finite and universal continuum limit at any fixed order in
perturbation theory and when the typical energies are much smaller than the
Landau pole. Therefore, when I refer to the continuum limit of QED, I always
have in mind some arbitrarily high, albeit finite, order in the perturbative
expansion. The existence and universality of the continuum limit in this weak
sense is a consequence of renormalizability by power counting, and of the fact
that the theory is weakly coupled at low energies. } In favour of QED$_\text{L}$
one may argue that none of the above prescriptions is completely free of problems
or potential subtleties. This does not come as a surprise as the constraint that
we are trying to circumvent (Gauss law) is not just a technical one, but has a
deep physical origin rooted in the very fundamental structure of QED. Reality is
that the subtleties arising in local prescriptions as QED$_\text{m}$ and
QED$_\text{C}$ can be systematically studied and accounted for with the powerful
tools of QFT. On the other hand no general tool exists for non-local
prescriptions and every solution to any newly found problem is a patch whose
validity may be debatable until such general tools are developed. My personal
recommendation is to use and further develop theoretically sound setups, as the
only way to eliminate unwanted, uncontrollable and unnecessary systematic errors
in calculations which aim the the percent precision.

\section{Charged states in a finite box: discussion of proposed prescriptions}
\label{sec:charged}

Gauss law forbids states with total electric charge different from zero in a box
with periodic boundary conditions. In fact the total electric charge is related
via the Gauss law to the integral of the divergence of the electric field, which
vanishes because of the boundary conditions
\begin{gather}
   Q = \int_{\mathbb{T}_3} \d^3 x \, j_0(t,\vec{x}) = \int_{\mathbb{T}_3} \d^3 x \, \partial_k E_k(t,\vec{x}) = 0 \ .
\end{gather}
This is obvious in classical electrodynamics. When temporal-gauge quantization
is considered (see e.g.~\cite{Creutz:1978xw}), the above equations is valid at
the operatorial level when restricted to the physical Hilbert subspace of
gauge-invariant states. The absence of charged states in the physical Hilbert
space is an essential obstruction for numerical simulations which aim at
calculating properties of charged particles.

A different way to see that charged states do not propagate in finite volume
with periodic boundary conditions is the following. Consider QED in a Euclidean
box with size $L_0 \times L_1  \times L_2 \times L_3$ and periodic boundary
conditions for all fields, and with with covariant gauge-fixing,
\begin{gather}
   S(A,\psi,\bar{\psi}) = \int \d^4 x \, \left\{ \frac{1}{4} F_{\mu\nu}^2 + \frac{1}{2\xi} ( \partial_\mu A_\mu )^2 + \bar{\psi} (\gamma_\mu D_\mu + m) \psi \right\} \ , 
   \label{eq:gen:action}
   \\
   D_\mu = \partial_\mu + i e A_\mu \ .
\end{gather}
The path-integral measure
\begin{gather}
   [ \d A ] \, [ \d \psi ] \, [ \d \bar{\psi} ] \ e^{-S(A,\psi,\bar{\psi})}
\end{gather}
is invariant under large gauge transformations
\begin{gather}
   A_\mu(x) \to A_\mu(x) - \frac{2 \pi n_\mu}{e L_\mu} \ , \nonumber \\
   \psi(x) \to e^{2\pi i (L^{-1}n)_\mu x_\mu} \psi(x) \ , \nonumber \\
   \bar{\psi}(x) \to e^{- 2\pi i (L^{-1}n)_\mu x_\mu} \bar{\psi}(x) \ .
   \label{eq:gen:large-def}
\end{gather}
Since large gauge transformations are not continuously connected to the identity, they survive any local gauge-fixing procedure. The product
$\psi(x) \bar{\psi}(y)$ transforms non-trivially under large gauge
transformations if $x \neq y$, therefore the two-point function of charged
fields vanishes at distinct points, i.e.
\begin{gather}
   \langle \psi(x) \bar{\psi}(y) \rangle = 0 \ , \quad \text{for } x \neq y \ .
\end{gather}
This shows that charged states do not propagate in finite volume with periodic boundary
conditions. All
prescriptions listed in section~\ref{sec:intro} solve this problem by destroying
large gauge transformations which prevent a non-zero charged two-point function
to be defined. The next subsections are devoted to reviewing particular aspects
and issues for each prescriptions.

Before moving to that, I want to add some remarks concerning the expectation
values generated by the QED action~\eqref{eq:gen:action}, which will be useful
later on. I will denote these expectation values simply by $\langle \star
\rangle$, defined formally as
\begin{gather}
   \langle P(A,\psi,\bar{\psi}) \rangle =
   \frac{
   \int [ \d A ] \, [ \d \psi ] \, [ \d \bar{\psi} ] \ e^{-S(A,\psi,\bar{\psi})} \ P(A,\psi,\bar{\psi})
   }{
   \int [ \d A ] \, [ \d \psi ] \, [ \d \bar{\psi} ] \ e^{-S(A,\psi,\bar{\psi})}
   }
   \ .
   \label{eq:gen:ex-val-def}
\end{gather}
This expression is only formal as both numerator and denominator contain a
multiplicative countable infinity due to large gauge transformations, which can
be trivially simplified as follows. Decompose the photon field into its constant
mode (a.k.a. global zero-mode) plus a fluctuation field
\begin{gather}
   A_\mu(x) = e^{-1} (L^{-1}\alpha)_\mu + B_\mu(x) \ , \nonumber \\
   (L^{-1}\alpha)_\mu \overset{\text{def}}{=} L_\mu^{-1} \alpha_\mu \overset{\text{def}}{=} \frac{e}{V} \int \d^4 x \, A_\mu(x) \ .
   \label{eq:gen:decomposition}
\end{gather}
The integration measure over the photon field can be factorized into the
integration measure over the constant mode times the integration measure over
the fluctuation. Consider an observable that transforms under an irreducible
representation of the (Abelian) group of large gauge
transformations~\eqref{eq:gen:large-def}, i.e.
\begin{gather}
   P_z(A,\psi,\bar{\psi}) \to  P_z(A,\psi,\bar{\psi}) \ \prod_\mu z_\mu^{n_\mu}
   \ ,
\end{gather}
for some $z \in \mathbb{C}^4$ and any $n \in \mathbb{Z}^4$. If $P_z$ transforms
non-trivially under large gauge transformations, i.e. $z \neq (1,1,1,1)$, then
its expectation value vanishes. If $P_z$ is invariant under large gauge
transformations, then the integration over the constant mode $\alpha$ can be
folded into the fundamental domain $(-\pi,\pi)^4$. In formulae
\begin{gather}
   \langle P_z(A,\psi,\bar{\psi}) \rangle =
   \begin{dcases}
      \frac{
      \int_{(-\pi,\pi)^4} \d^4 \alpha \int [ \d B ] \, [ \d \psi ] \, [ \d \bar{\psi} ] \ e^{-S(A,\psi,\bar{\psi})} \ P_z(A,\psi,\bar{\psi})
      }{
      \int_{(-\pi,\pi)^4} \d^4 \alpha \int [ \d B ] \, [ \d \psi ] \, [ \d \bar{\psi} ] \ e^{-S(A,\psi,\bar{\psi})}
      } \ ,
      \quad & \text{if } z = (1,1,1,1) \ ,
      \\
      0 \ , & \text{if } z \neq (1,1,1,1) \ ,
   \end{dcases}
   \label{eq:gen:folded-ex-val}
\end{gather}
where it is understood that, up to an immaterial normalization,
\begin{gather}
   [ \d B ] \overset{\text{def}}{=} \left[ \prod_x \d B(x) \right] \delta\left(\int \d^4 x \ B(x) \right) \ .
   \label{eq:gen:dB-def}
\end{gather}

\subsection{Absence of a transfer matrix in $\text{QED}_\text{TL}$}

For simplicity we consider periodic boundary conditions for fermions in all
directions. $\text{QED}_\text{TL}$ is simply defined by constraining the
constant mode of the gauge field to zero, i.e.
\begin{gather}
   \langle P(A,\psi,\bar{\psi}) \rangle_{\text{TL}} \overset{\text{def}}{=} \frac{1}{Z_\text{TL}} \int [\d A] \, [ \d \psi ] \, [ \d \bar{\psi} ] \ \delta\left(\int \d^4 x \ A(x) \right) \ e^{-S(A,\psi,\bar{\psi})} \ P(A,\psi,\bar{\psi})
   \ ,
\end{gather}
where $Z_\text{TL}$ is defined in such a way that $\langle 1 \rangle_\text{TL} =
1$. The insertion of a constraint that is non-local in time spoils the
transfer-matrix (i.e. Hamiltonian) representation of the path integral.

As noticed by the BMW collaboration in~\cite{Borsanyi:2014jba}, effective masses
for charged states defined with the $\text{QED}_\text{TL}$ prescription at fixed
order in $e^2$ diverge in the $T \to \infty$ limit. Let us be more precise, and
consider the two-point function
\begin{gather}
   C(t) \overset{\text{def}}{=} \frac{1}{2} \tr \frac{1+\gamma_0}{2} \int \d^3 x \ \langle \psi(t,\vec{x}) \bar{\psi}(0) \rangle_{\text{TL}}
   \label{eq:QEDTL:C} \ .
\end{gather}
The polarization projection is not essential and is introduced for the sole
purpose to simplify the formulae as much as possible. Explicit
calculation (see appendix~\ref{app:QEDTL}) of the effective mass shows that
\begin{gather}
   m_\text{eff}(t) \overset{\text{def}}{=} - \frac{\d}{\d t} \ln C(t) = m + \frac{e^2 T}{16m L^3} + O(e^2 T^0) + O(e^4) \ .
   \label{eq:QEDTL:divmeff}
\end{gather}
Notice that the linear term in $T$ vanishes in the $L\to \infty$ limit, showing
that the sickness of the prescription manifests itself as a non-commutation of
limits. Neglecting the fundamental issue that the $\text{QED}_\text{TL}$
prescription does not define a local QFT in finite volume, it has been suggested
\cite{Fodor:2016bgu} that it is enough to subtract the linear term in $T$ from
the effective masses in order to get meaningful results. However notice that, if
the $O(e^4)$ terms have a greater degree of divergence in $T$, these might
contribute with a non-negligible shift especially in full simulation at
unphysically large values of $e$. Whether the higher orders are under control or
not is completely unexplored territory.

At this point I would like to clarify a conceptual point. The fact that the
$O(e^2)$ effective mass is not finite in the $T \to \infty$ limit is certainly
an unappealing feature. However, contrarily to what stated by the BMW
collaboration, this fact alone does not imply the inexistence of the
Hamiltonian. If a Hamiltonian that is bounded from below exists then it is easy
to prove that the effective mass must be finite in the $T \to \infty$ limit. If
the Hamiltonian has a regular perturbative expansion (i.e. eigenvectors and
eigenvalues are infinitely differentiable with respect to the coupling in zero),
then it is also possible to prove that the effective mass must have a finite $T
\to \infty$ limit order by order in the perturbative expansion. However an IR
divergent effective mass (in the sense of the $T \to \infty$ limit) is not
incompatible with the existence of the Hamiltonian, as long as this does not
have a regular perturbative expansion. A simple example of this phenomenon based
on the harmonic oscillator is given in appendix~\ref{app:harmo}.

\subsection{Absence of a transfer matrix in $\text{QED}_\text{SF}$}

For simplicity we consider periodic boundary conditions for fermions in all
directions. $\text{QED}_\text{SF}$ is simply defined by means of
\begin{gather}
   \langle P(A,\psi,\bar{\psi}) \rangle_\text{SF} \overset{\text{def}}{=}
   \frac{
   \int_{(-\pi,\pi)^4} \d^4 \alpha \int [ \d B ] \, [ \d \psi ] \, [ \d \bar{\psi} ] \ e^{-S(A,\psi,\bar{\psi})} \ P(A,\psi,\bar{\psi})
   }{
   \int_{(-\pi,\pi)^4} \d^4 \alpha \int [ \d B ] \, [ \d \psi ] \, [ \d \bar{\psi} ] \ e^{-S(A,\psi,\bar{\psi})}
   }
   \label{eq:SF:def}
\end{gather}
for all observables. Notice that eq.~\eqref{eq:SF:def} coincides with
eq.~\eqref{eq:gen:folded-ex-val} for observables that are invariant under large
gauge transformations. Even though the restriction in the path integral in
eq.~\eqref{eq:SF:def} is non local, in the sector that is invariant under large
gauge transformations one can use the symmetry to undo the restriction and go
back to an expectation value defined in terms of a local action. One might
wonder whether the same happens for the two-point function of the charged
fermion. Let us see how it works.

Consider the product $\psi(x) \bar{\psi}(0)$, and let us introduce the following
non-local function of the photon field
\begin{gather}
   \mathcal{N}_\mu(A) = \left\lfloor \frac{\alpha_\mu+\pi}{2\pi} \right\rfloor
   =
   \left\lfloor \frac{1}{2} + \frac{e L_\mu}{2\pi V} \int \d^4 x \, A_\mu(x) \right\rfloor
   \ ,
\end{gather}
where $\lfloor x \rfloor$ denote the floor of $x$, i.e. the largest integer $n$
such that $n \le x$. Notice that $\mathcal{N}_\mu(A)=0$ when inserted in a
$\text{QED}_\text{SF}$ expectation value because of the restriction $-\pi <
\alpha_\mu < \pi$, therefore
\begin{gather}
   \langle \psi(x) \bar{\psi}(0) \rangle_\text{SF} = \langle e^{2 \pi i (L^{-1} x)_\mu \mathcal{N}_\mu(A)} \psi(x) \bar{\psi}(0) \rangle_\text{SF}
   \ .
\end{gather}
It is easy to check that, under a large gauge
transformation~\eqref{eq:gen:large-def},
\begin{gather}
   \mathcal{N}_\mu(A) \to \mathcal{N}_\mu(A) - n_\mu \ ,
\end{gather}
and the observable $e^{2 \pi i (L^{-1} x)_\mu \mathcal{N}_\mu(A)} \psi(x)
\bar{\psi}(0)$ is invariant. Since the $\text{QED}_\text{SF}$ expectation value
coincides with the unconstrained one for observables that are invariant under
large gauge transformations, one concludes that
\begin{gather}
   \langle \psi(x) \bar{\psi}(0) \rangle_\text{SF} = \langle e^{2 \pi i (L^{-1} x)_\mu \mathcal{N}_\mu(A)} \psi(x) \bar{\psi}(0) \rangle
   \ .
   \label{eq:QEDSF:2pf}
\end{gather}
Notice that eq.~\eqref{eq:QEDSF:2pf} shifts the non-locality from the
path-integral measure (in the $\text{QED}_\text{SF}$ expectation value) to the
observable (in the QED expectation value). In particular a Hamiltonian
representation for the two-point function seems not to be possible as, in the
formulation with a local path-integral measure, the observable is non-local in
time, moreover the non-locality range is equal to the whole lattice
size.\footnote{This is by no means a proof that a Hamiltonian representation
does not exist. In principle locality might be hidden by an inconvenient choice
of fundamental variables, and made manifest by a smart change of variables. I
have been privately informed by Schierholz that he is working on a proof that a
Hamiltonian representation of the two-point function actually exists, and this
would be an extremely interesting result.}

\subsection{(Non)renormalization of composite operators and $\text{QED}_\text{L}$}

The $\text{QED}_\text{L}$ prescription is defined by replacing the photon-field
integration measure with
\begin{gather}
   [ \d A ] \to [ \d A ] \ \left[ \prod_t \right] \, \prod_{\mu=0}^4 \delta\left( \int \d^3 x \, A_\mu(t,\vec{x}) \right) \ .
   \label{eq:QEDL:constr}
\end{gather}
This constraint is local in time, therefore $\text{QED}_\text{L}$ admits a
Hamiltonian. However this Hamiltonian is non-local in space. Locality is a core
property of quantum field theory, and several properties that we usually give
for granted depend on locality, such as renormalizability, operator product
expansion, effective-field theory description of low-energy modes. As we will
see, the renormalization of composite operators is broken by the constraint
defining $\text{QED}_\text{L}$. Before moving on, I want to comment on a couple
of myths existing around $\text{QED}_\text{L}$.
\begin{itemize}
   \item \textit{Since we are subtracting low-modes, i.e. IR modes, these should
   not affect the UV properties of the theory.} This is misleading on several
   levels. First one should notice that any notion of IR/UV decoupling (which is
   deeply related to the concept of effective field theory) makes sense only in
   a local theory. Second one should notice that the constraint in
   eq.~\eqref{eq:QEDL:constr} eliminates Fourier components with
   $p=(p_0,\vec{0})$ for any value of $p_0$ admitted by the boundary conditions.
   Therefore the constraint in eq.~\eqref{eq:QEDL:constr} eliminates momenta
   with arbitrarily large norm.
   
   \item \textit{The $\text{QED}_\text{L}$ prescription is equivalent to
   coupling $\text{QED}$ to a classical uniform charge density.} Let
   $\mathcal{J}_\mu(x)$ be a classical electromagnetic current. The action of
   $\text{QED}$ coupled to the classical current is
   \begin{gather}
      S_{\mathcal{J}} = S_0 + i e \int \d^4 x \ A_\mu(x) \mathcal{J}_\mu(x) \ .
   \end{gather}
   The dynamics of the quantum fields are determined by the probability
   distribution that defines the path-integral, while the dynamics of the
   classical current is determined by classical equations of motion. In
   particular the classical current should not be integrated over. The dynamics
   of the quantum fields in presence of a given classical current is local. $\text{QED}_\text{L}$ is obtained by
   choosing $\mathcal{J}_\mu(x)$ to be invariant under spatial translations, and
   by integrating over it in the path integral. The integration spoils the
   interpretation of $\mathcal{J}_\mu(x)$ as a classical current.
\end{itemize}

Let us discuss the renormalization of composite operators. For sake of
presentation, in place of $\text{QED}_\text{L}$, we shall consider the $\phi^4$
theory on $\mathbb{R} \times T_3$ where $T_3$ is the maximally symmetric
three-torus with linear size equal to $L$. We write the action having in mind
the renormalized perturbative expansion
\begin{gather}
   S = \int_{L^3} \d^4 x \left\{
   \frac{1}{2} \phi ( -\Box + m^2 ) \phi + \frac{\lambda}{4!} \phi^4
   + \frac{\delta Z}{2} \phi ( -\Box + m^2 ) \phi + \frac{\delta m^2}{2} \phi^2 + \frac{\delta \lambda}{4!} \phi^4
   \right\} \ ,
\end{gather}
where $\phi$ is the renormalized field, $m$ and $\lambda$ are renormalized
parameters, while $\delta Z$, $\delta m^2$ and $\delta \lambda$ are the
coefficients of the counterterms that are supposed to be at least $O(\lambda)$. We impose
a constraint on the path-integral measure that mimics $\text{QED}_\text{L}$
\begin{gather}
   [ \d \phi ] \to [ \d \phi ] \ \left[ \prod_t \right] \, \delta\left( \int \d^3 x \, \phi(t,\vec{x}) \right) \ .
   \label{eq:QEDL:phi4:constr}
\end{gather}
The propagator $\Delta'(x)$ in coordinate space can be split in two pieces
\begin{gather}
   \Delta'(x) = \Delta(x) - \Delta_0(x_0) \ ,
   \label{eq:QEDL:phi4:prop}
\end{gather}
the propagator in the full theory without constraint
\begin{gather}
   \Delta(x) = \frac{1}{L^3} \sum_{\vec{p} \in \frac{2\pi}{L} \mathbb{Z}^3} \int \frac{\d p_0}{2\pi}
   \frac{e^{ipx} e^{-\Lambda^{-2} (p^2+m^2)}}{p^2+m^2}
   \ ,
\end{gather}
and the subtraction
\begin{gather}
   \Delta_0(x_0) = \frac{1}{L^3} \int \frac{\d p_0}{2\pi}
   \frac{e^{ip_0x_0} e^{-\Lambda^{-2} (p_0^2+m^2)}}{p_0^2+m^2}
   \ ,
\end{gather}
where the heat-kernel regularization has been used. The counterterms $\delta Z$
and $\delta m$ are determined (up to a finite part) by requiring that the
two-point function $\langle \phi(x) \phi(0) \rangle$ is finite in the $\Lambda
\to \infty$ limit so long as $x \neq 0$. Analogously $\delta \lambda$ is
determined by requiring that the four-point function of the renormalized fields
is finite. The calculation of the action counterterms is standard, as the
constraint turns out to play no role in the identification of the divergences of
two and four-point functions at one loop. We give the results
\begin{gather}
   \delta Z = O(\lambda^2)
   \ , \quad
   \delta m^2 = - \frac{\lambda \Lambda^2}{2(4\pi)^2} + \frac{\lambda m^2}{(4\pi)^2} \ln \frac{\Lambda}{\mu} + O(\lambda^2)
   \ , \quad
   \delta \lambda = \frac{3 \lambda^2}{(4\pi)^2} \ln \frac{\Lambda}{\mu} + O(\lambda^3)
   \ .
\end{gather}

As a consequence of the constraint in eq.~\eqref{eq:QEDL:phi4:constr}, the
propagator $\Delta'(x)$ is not a smooth function for $x \neq 0$. In fact a
little algebra shows that
\begin{gather}
   \Box \Delta'(0,\vec{x})
   =
   -\frac{\Lambda}{(4\pi)^{1/2} L^3} + O(\Lambda^0)
   \label{eq:QEDL:phi4:basicdiv}
\end{gather}
for any $\vec{x} \neq \vec{0}$. This non-local divergence will propagate into
loops by generating unwanted mixing with non-local operators. Let us see how in a
particular example, which has been cooked up to show this feature in a very
simple diagram.

We consider the bare operator $(\Box \phi)^2$. In a local theory, the
divergences generated by the insertion of $(\Box \phi)^2(x)$ in general
expectation values can be subtracted by inserting local counterterms in $x$.
These counterterms are all possible local operators with the same quantum
numbers as $(\Box \phi)^2$, and with dimension not greater than $6$. In other
terms, a renormalized operator can be defined as follows
\begin{gather}
   [ (\Box \phi)^2 ]_R(x) = (\Box \phi)^2(x) + \sum_{d_\mathcal{O} \le 6} c_\alpha \Lambda^{6-d_\mathcal{O}} \mathcal{O}(x) + c_\text{id} \Lambda^6 \ ,
   \label{QEDL:phi4:OPE}
\end{gather}
where the $c_\alpha$'s are $O(\lambda)$ coefficients which diverge at most
logarithmically ($c_\text{id}$ is the so called mixing with the identity). The
renormalized operator is defined such that all expectation values of the form
\begin{gather}
   \langle \phi(z^1) \cdots \phi(z^n) [ (\Box \phi)^2 ]_R(x) \rangle
\end{gather}
are finite in the $\Lambda \to \infty$ limit so long as all point are pairwise
distinct. We will show explicitly that in the theory with
constraint~\eqref{eq:QEDL:phi4:constr}, it is not possible to choose local
counterterms as in eq.~\eqref{QEDL:phi4:OPE} so that the following expectation
value is finite
\begin{gather}
   \langle \phi(z^1) \phi(z^2) [ (\Box \phi)^2 ]_R(x) \rangle_c \ .
\end{gather}
The connected expectation value is taken so to kill the mixing with the
identity. We can calculate
\begin{gather}
   \langle \phi(z^1) \phi(z^2) (\Box \phi)^2(x) \rangle_c
   =
   2 \langle \phi(z^1) \Box \phi(x) \rangle \, \langle \phi(z^2) \Box \phi(x) \rangle
   + \nonumber \\ \qquad
   - \lambda \int \d^4 y \ \Delta'(z^1-y) \Delta'(z^2-y) [\Box \Delta'(x-y)]^2 + O(\lambda^2)
   \ .
\end{gather}
The two-point functions in the r.h.s. have the same divergence as in
eq.~\eqref{eq:QEDL:phi4:basicdiv} when $z^1_0=x_0$ for all values of
$\vec{z}^1$, or $z^2_0=x_0$ for all values of $\vec{z}^2$. This divergence is
obviously non-local. Even when this divergence is avoided, the loop integral has
extra divergences with respect to the infinite-volume case. The full analysis of
divergences is quite tedious but straightforward. One can expand the propagators
$\Delta'(x-y)$ inside the loop by means of eq.~\eqref{eq:QEDL:phi4:prop}. It is
interesting to look at one particular contribution, i.e. the one that is
obtained by replacing both loop propagators by $-\Delta_0(x_0-y_0)$. Some
calculation yields
\begin{gather}
   - \lambda \int \d^4 y \ \Delta'(z^1-y) \Delta'(z^2-y) [\partial_0^2 \Delta_0(x_0-y_0)]^2
   = \nonumber \\=
   - \frac{\lambda}{(8\pi)^{1/2}L^6} \Lambda \int \d^4 y \ \Delta'(z^1-y) \Delta'(z^2-y) \, \delta(y_0-x_0)
   + O(\Lambda^0)
   \label{eq:QEDL:phi4:lineardiv}
   \ .
\end{gather}
The appearance of this divergence is fairly simple to understand. If we set
$\Lambda=\infty$ then by definition,
\begin{gather}
   \partial_0^2 \Delta_0(x_0-y_0) = - \delta(x_0-y_0) + m^2 \Delta_0(x_0-y_0) \ ,
\end{gather}
and when we square this expression, we get a $[\delta(x_0-y_0)]^2$ which is a
linear divergence, whose exact value can be calculated only at finite regulator.
The divergence in eq.~\eqref{eq:QEDL:phi4:lineardiv} would be canceled by the
insertion in the r.h.s. of eq.~\eqref{QEDL:phi4:OPE} of the following operator
\begin{gather}
   \frac{\lambda}{2(8\pi)^{1/2}L^6} \Lambda \int \d^3 \vec{y} \ \phi^2(x_0,\vec{y}) \ ,
\end{gather}
which is obviously non local. Again the effect of non-locality shows as a
non-commutation of limits. If we take the infinite volume limit before the
$\Lambda \to \infty$ limit, this contribution vanishes.

Even though renormalization at one loop of operators with dimension not greater
than 4 seems to happen in the same way as in infinite volume, I have shown that
renormalization by local counterterms breaks down even at one loop for operators
with high enough dimension. This will obviously propagate into the Symanzik
expansion of low dimensional operators (and in particular of the action) with
unexplored consequences.

Translating the above example to $\text{QED}_\text{L}$ is not trivial, as the
photon propagator cannot be decomposed in a simple way as in
eq.~\eqref{eq:QEDL:phi4:prop}. Also one may need to go to higher order to
reproduce an analogous mechanism. A full analysis of the finite-volume
divergences in $\text{QED}_\text{L}$ is nowhere in sight, and I hope that I have
convinced the reader that this is not a trivial matter. The above analysis is
maximally relevant for strategies that rely on unphysically large values of
$\alpha_\text{em}$ as higher-loop effects are amplified, and on simultaneous
extrapolations to $L \to \infty$ and $a \to 0$ as these two limits do not
commute. Precision can not be claimed without having the effects of non-locality
under control.

\subsection{A local formulation: $\text{QED}_\text{m}$}

$\text{QED}_\text{m}$ is defined by giving a mass to the photon, i.e. by
replacing the action with
\begin{gather}
   S_\text{m}(A,\psi,\bar{\psi}) \overset{\text{def}}{=} S(A,\psi,\bar{\psi}) + \frac{m_\gamma^2}{2} \int \d^4x \ A_\mu^2 \ ,
   \label{eq:QEDm:action}
\end{gather}
which gives a consistent (in particular, local) QFT in finite
volume, defined order by order in the perturbative expansion. The mass term
introduces a soft breaking of gauge symmetry (like the gauge-fixing term). The
renormalization of $\text{QED}_\text{m}$ is well understood. The $m_\gamma \to
0$ limit in infinite volume is widely studied in the literature. However
$\text{QED}_\text{m}$ employs two IR regulators: the photon mass and the finite
volume. The $L \to \infty$ and $m_\gamma \to 0$ limits do not commute, and
infinite-volume QED is recovered only if the $L \to \infty$ limit is taken
before the $m_\gamma \to 0$ limit. 
Non-commutation of limits makes usually the extrapolation particularly
challenging. However it has to be noted that, as long as $m_\gamma \neq 0$, the
infinite-volume limit is reached easily as the finite-volume corrections are
exponentially small (e.g. for masses of stable states). I want to investigate
here a bit more in detail the implications of the non-commutation of limits.

The lengthy calculation in appendix~\ref{app:QEDm} shows that the fermion
two-point function in $\text{QED}_\text{m}$ can be represented as follows
\begin{gather}
   \langle \psi(x) \bar{\psi}(0^+) \rangle_\text{m}
   =
   \frac{
   \sum_{q \in \mathbb{Z}^4} e^{- \frac{e^2}{2 m_\gamma^2 V} (L q+x)_\mu^2}
   \langle \psi(x) \bar{\psi}(0^+)
   e^{- \frac{m_\gamma^2}{2} \int \d^4 z \ B^2_\mu(z)} \delta_{Q,q} \rangle_\text{TL}
   }{
   \sum_{q \in \mathbb{Z}^4} e^{- \frac{e^2}{2 m_\gamma^2 V} (L q)_\mu^2}
   \langle e^{- \frac{m_\gamma^2}{2} \int \d^4 z \ B^2_\mu(z)} \delta_{Q,q} \rangle_\text{TL}
   }
   \ ,
   \label{eq:QEDm:2pt-2}
\end{gather}
where $Q_\mu$ is the electric charge (in units of $e$) operator defined by
interpreting $\mu$ as temporal direction and localized on the time-slice
$x_\mu=0$, i.e.
\begin{gather}
   Q_\mu = \int \d^4 z \ \delta(z_\mu) \bar{\psi} \gamma_\mu \psi(z) \ .
\end{gather}

We want to study the $m_\gamma \to 0$ limit at fixed volume of the two-point
function in time momentum representation
\begin{gather}
   C(t,\vec{p}) \overset{\text{def}}{=} \int_{L_1 \times L_2 \times L_3} \d^3 x \ e^{-i \vec{p}\vec{x}} \, \langle \psi(t,\vec{x}) \bar{\psi}(0^+) \rangle_\text{m} \ .
   \label{eq:QEDm:2pt-3}
\end{gather}
In the $m_\gamma \to 0$ limit, the denominator of eq.~\eqref{eq:QEDm:2pt-2} is
dominated by the term $q=0$, i.e.
\begin{gather}
   \sum_{q \in \mathbb{Z}^4} e^{- \frac{e^2}{2 m_\gamma^2 V} (L q)_\mu^2}
   \langle e^{- \frac{m_\gamma^2}{2} \int \d^4 z \ B^2_\mu(z)} \delta_{Q,q} \rangle_\text{TL}
   \overset{m_\gamma \to 0}{=}
   \langle \delta_{Q,0} \rangle_\text{TL} + O(m_\gamma^2)
   \ .
\end{gather}
When the numerator of eq.~\eqref{eq:QEDm:2pt-2} is plugged into
eq.~\eqref{eq:QEDm:2pt-3}, the $m_\gamma \to 0$ limit is obtained by saddle
point. For $0<t<L_0/2$ the saddle point is located at $q=0$ and $\vec{x}=0$,
while for $L_0/2 < t < L_0$ the saddle point is located at $q=(-1,\vec{0})$ and
$\vec{x}=0$. Let us consider only the $0<t<L_0/2$ case for simplicity, i.e.
\begin{gather}
   C(t,\vec{p})
   \overset{m_\gamma \to 0}{=}
   \frac{(2\pi e^{-2} m_\gamma^2 V)^{3/2}}{
   \langle \delta_{Q,0} \rangle_\text{TL}
   }
   e^{- \frac{e^2}{2 m_\gamma^2 V} t^2}
   \langle \psi(t,\vec{0}) \bar{\psi}(0^+) \delta_{Q,0} \rangle_\text{TL}
   \{ 1 + O(m_\gamma^2) \}
   \ .
   \label{eq:QEDm:2pt-4}
\end{gather}
The two-point function vanishes exponentially in the $m_\gamma \to 0$ limit for
$t \neq 0$ as it should, since large gauge transformations become symmetries in
the $m_\gamma \to 0$ limit. Moreover the leading term in the $m_\gamma \to 0$
limit does not depend on the momentum $\vec{p}$, as a direct consequence of the
fact that the integral is dominated by the value of the integrand at $\vec{x} =
\vec{0}$. At this order, the effective mass is
\begin{gather}
   m_\text{eff}(t,\vec{p}) \overset{\text{def}}{=} - \frac{\d}{\d t} \ln C(t,\vec{p}) = 
   \frac{e^2}{m_\gamma^2 V} t - \frac{\d}{\d t} \ln \langle \psi(t,\vec{0}) \bar{\psi}(0^+) \delta_{Q,0} \rangle_\text{TL} + O(m_\gamma^2)
   \ .
   \label{eq:QEDm:2pt-5}
\end{gather}
The term proportional to $t$ in the effective mass is the one identified
in~\cite{Endres:2015gda}. Notice that this term vanishes in the infinite-volume
limit and diverges in the $m_\gamma \to 0$ limit. Even if this term is removed
by hand, the non-commutation of limits is not resolved. If $m_\gamma$ is too
small, the physical dependence of the effective mass on the momentum is
suppressed, and effectively one is just using a complicated method to extract
the $\text{QED}_\text{TL}$ effective mass in coordinate space. In this regime it
would be impossible to construct states with definite momentum, which are
necessary e.g. in the calculation of transition amplitudes.
It is worth noticing that the term proportional to $t$ is large in the exploratory study in~\cite{Endres:2015gda}, which suggests in fact that this study is performed in the regime in which eq.~\eqref{eq:QEDm:2pt-5} is valid. A possible way to verify or disprove this is to calculate the effective mass at different values of $\vec{p}$ and check whether it is constant or not.

\subsection{A local formulation: $\text{QED}_\text{C}$}

$\text{QED}_\text{C}$ is defined by means of C$^\star$ boundary conditions
(a.k.a. C-parity boundary conditions) along the spatial directions for all
fields,
\begin{gather}
   A_\mu(x+ L_k \hat{k}) = -  A_\mu(x) \ , \\
   \psi(x+ L_k \hat{k}) = C^{-1} \bar{\psi}^T(x) \ ,
\end{gather}
where $C$ is the charge-conjugation matrix for spinors. $\text{QED}_\text{C}$ is
a consistent (in particular, local) QFT at finite volume, defined order by order
in the perturbative expansion. Some features and delicate aspects of
$\text{QED}_\text{C}$ are listed in the following.

\begin{itemize}
   
   \item The total electric charge is not constrained to be zero by the Gauss
   law
   \begin{gather}
      Q = \int_{\mathbb{T}_3} \d^3 x \, j_0(t,\vec{x}) = \int_{\mathbb{T}_3} \d^3 x \, \partial_k E_k(t,\vec{x}) = -2 \sum_k \int \d^3 x \ \delta(x_k) \ E_k(t,\vec{x}) \ ,
   \end{gather}
   since the electric field is antiperiodic in the spatial directions. In other
   works, the electric flux produce by a particle can escape the torus and be
   eaten by the image particle (which has opposite charge).
   
   \item Large gauge transformations of the form~\eqref{eq:gen:large-def} do not
   leave the photon field boundary conditions inveriant and are therefore not a symmetry
   of the theory.
   
   \item Charge conservation and, similarly, flavor conservation are partially
   broken by the boundary conditions. The electric charge $Q$ is not conserved,
   but the quantum number $(-1)^Q$ is. The same happens for each individual
   flavor number. This is enough to protect all stable mesons and most of the
   stable baryons from unphysical decays. Some baryons, that would be stable in
   infinite volume, mix with lighter ones because of the boundary conditions
   (e.g. $\Xi^-$ mixes with $p$). In effective masses this effect appears
   (again!) as a non-commutation of IR limit
   \begin{gather}
      m_\text{eff}(t|L) = - \lim_{T \to \infty} \frac{\d}{\d t} \ln \int \d^3 x \ \langle \Xi_-^\dag(t,\vec{x}) \Xi_-(0,\vec{x}) \rangle
      \overset{\text{def}}{=} - \frac{\d}{\d t} \ln C(t|L)
      \ , \\
      \lim_{t \to \infty} \lim_{L \to \infty} m_\text{eff}(t|L) = M_{\Xi^-}
      \ , \qquad
      \lim_{L \to \infty} \lim_{t \to \infty} m_\text{eff}(t|L) = M_{p}
      \ .
   \end{gather}
   These spurious mixing are due to flavored mesons traveling around the torus
   and are suppressed exponentially with the volume, more precisely the
   following decomposition for the correlator $C(t|L)$ exists
   \begin{gather}
      C(t|L) = C_0(t|L) + e^{-\mu L} C_1(t|L) \ ,
   \end{gather}
   where both $C_0(t|L)$ and $C_1(t|L)$ are finite in the $L \to \infty$ limit
   at fixed $t$, and also
   \begin{gather}
      - \lim_{t \to \infty} \frac{\d}{\d t} \ln C_0(t|L) = M_{\Xi^-}(L)
      \ , \qquad
      - \lim_{t \to \infty} \frac{\d}{\d t} \ln C_1(t|L) = M_{p}(L) \ .
   \end{gather}
   The suppression exponent $e^{-\mu L}$ is calculated to be $O(10^{-10})$ at
   physical quark masses and $M_\pi L = 4$. 
   
   \item The partial breaking of flavor symmetry does not affect the
   renormalization of composite operators. Thanks to locality, the divergent
   part of the renormalization constants does not depend on the volume.
   
   \item The non-universal finite volume corrections to the hadron masses vanish
   as $L^{-4}$ (as opposed to $L^{-3}$ in QED$_\text{L}$).
   
   \item A gauge-invariant representation for the charged-state interpolating
   operators is possible.
   
\end{itemize}

\section{IR divergences and decay rates}
\label{sec:Smatrix}

In this section I will briefly discuss radiative corrections to decay rates.
Subsection~\ref{subsec:soft} is completely textbook, and contains a review of
some basic facts about IR soft divergences. I have chosen to include this
subsection since IR divergences do not belong to the standard toolbox of the
lattice community. Subsection~\ref{subsec:RM-SOTON} is a review of the
exploratory calculation of the pion and kaon decay rate at $O(\alpha_\text{em})$
presented at this conference by the RM-SOTON collaboration. This subsections
gives me the chance also to discuss the separation of radiative corrections in a
universal piece (a.k.a. \textit{inner-bremsstrahlung}) and a
\textit{structure-dependent} piece, which is a useful tool for numerical
simulations, and it is also fundamental in order to understand the jargon of the
experimentalist and the phenomenologist. Finally in
subsection~\ref{subsec:collinear} I argue that radiative corrections can be
enhanced by large logarithms due to quasi-collinear divergences in decay rates
of heavy mesons. This phenomenology is completely non-perturbative and hinders
the reliability of phenomenological estimates of radiative corrections. It is
where our community may be able to give an important contribution hopefully in
the near future.

\subsection{Some facts about soft divergences}
\label{subsec:soft}

If one wants to extract S-matrix elements from numerical simulations when QCD is
coupled to QED, besides dealing with the standard complications due to the
Euclidean setup and the finite volume, one has to deal also with IR divergences.
We assume that all charged asymptotic particles are massive. In this case IR
(\text{soft}) divergences are due to soft photons (i.e. photons with
four-momentum $k$, such that $k^2$ is asymptotically small) in loops and in the
final state.

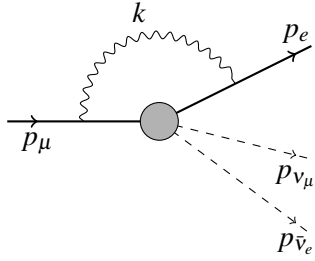
\begin{figure}
   
   \centering
   
   \tikzset{
     photon/.style={decorate, decoration={snake,amplitude=.5mm,segment length=1.8mm}, draw=black}
   }

   \begin{tikzpicture}
      
   \draw[thick,postaction={decorate},decoration={markings,mark=at position 0.2 with {\arrow{>}}}] (0,0) -- (2,0) node[pos=0.2,below] {$p_\mu$};
   \draw[thick,postaction={decorate},decoration={markings,mark=at position 0.9 with {\arrow{>}}}] (2,0) -- (4,1) node[pos=0.9,above] {$p_e$};
   \draw[dashed,postaction={decorate},decoration={markings,mark=at position 0.9 with {\arrow{>}}}] (2,0) -- (4,-0.5) node[pos=0.9,below] {$p_{\nu_\mu}$};
   \draw[dashed,postaction={decorate},decoration={markings,mark=at position 0.9 with {\arrow{>}}}] (2,0) -- (4,-1.5) node[pos=0.9,below] {$p_{\bar{\nu}_e}$};
   \draw[fill=black!30] (2,0) circle (.25);
   \draw[photon] (1,0) .. controls (1,1.2) and (2.3,1.5) .. (3,.5) node[pos=0.5,above=1mm] {$k$};

   \end{tikzpicture}
   
   \label{fig:mudecay}
   \caption{Feynman diagram that contributes to the $\mu^- \to e^- + \bar{\nu}_e +
   \nu_\mu$ amplitude at order $\alpha_\text{em}$. The gray circle represents the
   insertion of the EW effective Hamiltonian. The loop integral is IR
   divergent.}

\end{figure}

Let us review briefly how soft divergences are generated in loops. Roughly
speaking they are due to the fact that the photon propagator $1/k^2$ is singular
when $k^2 \to 0$. However it has to be noted that this singularity is
integrable, hence generally not enough to produce a divergence in the Feynman
integrals. In fact $n$-point functions in coordinate space (both in Minkowskian
and Euclidean space-time) are IR finite. Soft divergences appear in 1PI diagrams
when external momenta are chosen on the mass shell. In order to illustrate how
this happens, let us consider the EW decay $\mu^- \to e^- + \bar{\nu}_e +
\nu_\mu$. For simplicity we can work with Fermi's effective theory (i.e. in the
$M_W \to \infty$ limit), and consider the one-loop 1PI diagram in
figure~\ref{fig:mudecay}. The S-matrix element is calculated by setting the
external momenta on the mass-shell, i.e. $p_\mu^2=-m_\mu^2$ and $p_e^2=-m_e^2$
(in Euclidean notation). If $k$ is the photon momentum, the charged-particle
propagators inside the loop are
\begin{gather}
   \frac{-i (\fsl{p}_\star + \fsl{k}) + m_\star}{ (p_\star + k)^2 + m_\star^2 }
   =
   \frac{-i (\fsl{p}_\star + \fsl{k}) + m_\star}{ 2 p_\star k + k^2 }
   =
   \frac{-i \fsl{p}_\star + m_\star}{ 2 p_\star k } + O(k^0)
   \ .
\end{gather}
In the $k \to 0$ limit the charged
propagators go on-shell and contribute with a $k^{-1}$ singularity to the loop
integrand. Since there is two of them, and a photon propagator, the loop
integral restricted to the soft region is proportional to
\begin{gather}
   \int_{k^2 < \Lambda} \frac{\d^4 k}{(2\pi)^4} \frac{1}{ 2 p_\mu k } \frac{1}{ 2 p_e k } \frac{1}{k^2}
   \ ,
\end{gather}
which is logarithmically divergent. Notice that only one singular matter
propagator would not be enough to generate the divergence.

At fixed order in $\alpha_\text{em}$ transition amplitudes are IR divergent
because of loops of soft photons. Bloch and Nordsieck~\cite{Bloch:1937pw}
pointed out that these divergences cancel in decay rates, when integrated over
an arbitrary number of soft photons in the final state. Let us review briefly
the physics significance of the Bloch-Nordsieck cancellation mechanism and how
it works. Let us consider a generic process $\alpha \to \beta$ involving any
number of hard particles of any kind. Let $\vec{p}_n$ be the momentum of the
$n$-th particle in this process. Let us consider also the process $\alpha \to
\beta + N\gamma$ in which the final state contains also $N$ soft photons, with
the following properties
\begin{enumerate}[noitemsep,nolistsep]
   \item the $n$-th incoming particle has momentum $\vec{p}'_n=\vec{p}_n$;
   \item the $n$-th outgoing hard particle has momentum $\vec{p}'_n$, with $|\vec{p}'_n-\vec{p}_n|<\Delta p$;
   \item each soft outgoing photon has energy less than $\Delta E$.
\end{enumerate}
If $\Delta E$ and $\Delta p$ are the energy and momentum resolution of the
detector, then the two processes are experimentally indistinguishable. The
probability rate to transition from $\alpha$ to $\beta$ turns out to be infinite
at any order in $\alpha_\text{em}$ beyond tree-level. However this probability
rate is not relevant from the experimental point of view. In fact only the
probability rate to transition from $\alpha$ to any final state that is
experimentally indistinguishable from $\beta$ is relevant. It turns out that the
latter probability rate is IR finite at any order in the perturbative expansion.

Let us try to be a bit more precise.\footnote{
The upcoming discussion is textbook (see e.g. section 13.4 in~\cite{StefanoI}).
The concepts of \textit{factorization}, \textit{exponentiation} and
\textit{universality} of soft divergences have their root in the fundamental
work of Yennie, Frautschi, Suura~\cite{Yennie:1961ad}, and Grammer,
Yennie~\cite{Grammer:1973db}, and have been beautifully summarized by
Weinberg~\cite{Weinberg:1965nx}. The interpretation of IR divergences as
break-down of the perturbative expansion of transition amplitudes has been
analyzed in detail by Lee and Nauenberg~\cite{Lee:1964is}.
}
Let us introduce an IR regulator, e.g. a mass $m_\gamma$ for the photon. Let
$M_{\alpha \beta}(\vec{k}_1,\dots,\vec{k}_N;m_\gamma)$ be the transition
amplitude for the process $\alpha \to \beta + N \gamma$, where $\vec{k}_r$ is
the momentum the $r$-th soft photon. Notice that the soft photons cannot carry a
large amount of total energy away (even if $N$ is large and $m_\gamma=0$),
otherwise by energy conservation the energy of some of the outgoing particles
would be notably affected. Therefore it makes sense to consider the transition
rate summed over any number of outgoing soft photons with individual energy less
than $\Delta E$ and total energy less then $E_T=O(\Delta p)$, i.e.
\begin{gather}
   \label{eq:IR:transition-0}
   \Gamma_{\alpha\beta}(\Delta E,E_T;m_\gamma)
   = \\ \qquad =
   \mathcal{N}_\alpha \sum_{N=0}^\infty \frac{1}{N!}
   \int_{\substack{|\vec{k}_r|<\Delta E \\ \sum_{r=1}^N |\vec{k}_r| < E_T}}
   \left\{ \prod_{r=1}^N \frac{\d^3 k_r}{(2\pi)^3 2 (m_\gamma^2 + \vec{k}^2)^{1/2}} \right\}
   \left| M_{\alpha\beta}(\vec{k}_1,\dots,\vec{k}_N;m_\gamma) \right|^2
   \ ,  \nonumber
\end{gather}
where $\mathcal{N}_\alpha$ is a normalization factor depending only on the
initial state, and the sum over the polarizations of the soft photons is
understood. The nontrivial fact about soft divergences is that the asymptotic
behaviour of the transition amplitude in the $m_\gamma \to 0$ and $\vec{k} \to
0$ limit is calculable at all orders in the perturbative expansion
\begin{gather}
   \label{eq:IR:amplitude}
   M_{\beta\alpha}(\vec{k}_1,\dots,\vec{k}_N;m_\gamma)
   \overset{\vec{k} \sim m_\gamma \to 0}{\simeq}
   \hat{M}_{\beta\alpha}(\mu) \left( \frac{m_\gamma}{\mu} \right)^{\frac{\alpha_\text{em}}{2} A_{\beta\alpha}} 
   \prod_{r=1}^N \sum_n \frac{e q_n \eta_n \ p_n \cdot \epsilon_r^*}{p_n \cdot k_r}
   \ .
\end{gather}
Here $p_n$, $m_n$, and $q_n$ are the four-momentum, mass and electric charge (in
units of $e$) of the $n$-th hard particle, while $\eta_n$ is defined to be equal
to $+1$ and $-1$ for particles in the final and initial state respectively.
$k_r$ and $\epsilon_r$ are the four-momentum and polarization four-vector of the
$r$-th soft photon. $\mu$ is an arbitrary energy scale introduced to the sole
purpose to match the dimensions, $\hat{M}_{\beta\alpha}(\mu)$ does not depend on
$m_\gamma$. Finally \ $A_{\beta\alpha}$ is a known exponent
\begin{gather}
   A_{\beta\alpha} \overset{\text{def}}{=}
   - \frac{1}{2\pi} \sum_{nm} \frac{q_n q_m \eta_n \eta_m}{\beta_{nm}} \ln \frac{1+\beta_{nm}}{1-\beta_{nm}}
   + i \sum_{n \neq m} \frac{q_n q_m}{\beta_{nm}} \delta_{\eta_n,\eta_m}
   \ , \\
   \beta_{nm} \overset{\text{def}}{=} \sqrt{ 1 - \frac{m_n^2 m_m^2}{(p_n \cdot p_m)^2} }
   \ .
\end{gather}
It is usually said that soft divergences \textit{factorize}, since the virtual
soft photons contribute to the transition amplitude with the factor
$(m_\gamma/\mu)^{\frac{\alpha_\text{em}}{2} A_{\beta\alpha}}$, and each real
soft photon contributes with a factor $\sum_n e q_n \eta_n (p_n \cdot
\epsilon_r^*)/(p_n \cdot k_r)$. It is also usually said that soft divergences
are \textit{universal} in the sense that these factors are completely determined
by masses, charges and momenta of the particles in the initial and final
states. In particular the factors depend neither on the microscopic interactions
that produce the scattering process nor on the spin and internal structure of
the particles.

Notice that the virtual-photon factor gives contributions to all orders in the
perturbative expansion. The real part of $A_{\alpha\beta}$ is proven to be
always positive, therefore the formally resummed transition amplitude vanishes
when the IR regulator is removed while the momenta of the soft photons are kept
small but non-vanishing. This phenomenon is known as \textit{evaporation of the
S-matrix}.\footnote{
When expanded to a given order in $\alpha_\text{em}$, the exponential factor
\begin{gather}
   \left( \frac{m_\gamma}{\mu} \right)^{\frac{\alpha_\text{em}}{2} A_{\beta\alpha}}
   =
   \sum_{N=0}^\infty \frac{1}{N!} \left( \frac{\alpha_\text{em}}{2} A_{\beta\alpha} \ln \frac{m_\gamma}{\mu} \right)^N \ .
\end{gather}
generates logarithmic divergences. At fixed order in the perturbative expansion
the transition amplitude does not vanish in the $m_\gamma \to 0$ limit as in the
resummed case, on the contrary it diverges logarithmically. This phenomenon may
look counter-intuitive, but the mathematical mechanism behind it is quite
trivial and shows that the perturbative expansion is broken by large calculable
IR logarithms.
}
If $\Delta E$, $E_T$ and $m_\gamma$ are small enough, one can plug the
approximation~\eqref{eq:IR:amplitude} into the
formula~\eqref{eq:IR:transition-0} for the integrated transition rate.
Because of the singularity at $\vec{k}_r \to 0$, the phase-space integral in
eq.~\eqref{eq:IR:transition-0} diverges when the IR regulator is removed. This
divergence compensates exactly the vanishing factor due to virtual soft photons,
so that the transition probability rate becomes finite and non-zero in the
$m_\gamma \to 0$ limit,
\begin{gather}
   \Gamma_{\alpha\beta}(\Delta E,E_T) = \lim_{m_\gamma \to 0 } \Gamma_{\alpha\beta}(\Delta E,E_T;m_\gamma)
   \overset{\Delta E \sim E_T \to 0}{\simeq}
   \nonumber \\ \qquad \simeq
   \mathcal{N}_\alpha F(\tfrac{\Delta E}{E_T},\Re A_{\beta\alpha})
   \left( \frac{\Delta E}{\mu} \right)^{\alpha_\text{em} \Re A_{\beta\alpha}} |\hat{M}_{\beta\alpha}(\mu)|^2
   \ ,
   \label{eq:IR:transition-1}
\end{gather}
where $F(x,A)$ is a known kinematic function which is reported for completeness
\begin{gather}
   F(x,A) = \frac{1}{\pi} \int_{-\infty}^\infty \d u \frac{\sin u}{u} \exp \left( \alpha_\text{em} A \int_0^x \d \omega \frac{e^{i \omega u}-1}{\omega} \right)
   \ .
\end{gather}
Notice that the transition rate in the resummed form~\eqref{eq:IR:transition-1}
is finite for any value of $\Delta E$ and vanishes in the $\Delta E \to 0$
limit. However at fixed order in the perturbative expansion
\begin{gather}
   \Gamma_{\alpha\beta}(\Delta E,E_T)
   \overset{\Delta E \sim E_T \to 0}{\simeq}
   \mathcal{N}_\alpha F(\tfrac{\Delta E}{E_T},\Re A_{\beta\alpha}) \left( 1 + \alpha_\text{em} \Re A_{\beta\alpha} \ln \frac{\Delta E}{\mu} \right) |\hat{M}_{\beta\alpha}(\mu)|^2 + O(\alpha_\text{em}^2)
\end{gather}
the transition rate is logarithmically divergent in the $\Delta E \to 0$ limit.
This divergence is not real, but it signals the breakdown of the perturbative
expansion in processes with soft photons in the final state. If we consider for
instance the leptonic decay of the hadron $h^- \to \ell^- + \bar{\nu}_\ell$,
then one can calculate explicitly
\begin{gather}
   A_{\beta\alpha} = \frac{2}{\pi} \left\{ \frac{m_h^2+m_\ell^2}{m_h^2-m_\ell^2} \ln \frac{m_h}{m_\ell} - 1 \right\}
   = \begin{cases}
   2.9(1) \qquad & \text{for } \pi^- \to e^- + \bar{\nu}_e \\ 
   1.6(1) \times 10^{-2} \qquad & \text{for } \pi^- \to \mu^- + \bar{\nu}_\mu \\
   5.2(1) \qquad & \text{for } B^- \to e^- + \bar{\nu}_e \\ 
   1.9(1) \qquad & \text{for } B^- \to \mu^- + \bar{\nu}_\mu
   \end{cases}
   \ .
   \label{eq:IR:Ah}
\end{gather}
For a back-of-the-envelope calculation, one can take $\mu$ equal to the largest
scale in the problem, i.e. the hadron mass, and calculate $\overline{\Delta
E}$ with the property that for all $\Delta E < \overline{\Delta E}$ the
$O(\alpha_\text{em})$ corrections are larger than $10 \%$, i.e.
\begin{gather}
   \overline{\Delta E}
   \simeq \begin{cases}
   0.5 \text{ MeV} \qquad & \text{for } \pi^- \to e^- + \bar{\nu}_e \\ 
   10^{-362} \text{ MeV} \qquad & \text{for } \pi^- \to \mu^- + \bar{\nu}_\mu \\
   400 \text{ MeV} \qquad & \text{for } B^- \to e^- + \bar{\nu}_e \\ 
   3 \text{ MeV} \qquad & \text{for } B^- \to \mu^- + \bar{\nu}_\mu
   \end{cases}
   \ .
\end{gather}
Notice that this energy varies a lot with the process. It is obvious that
higher-order radiative corrections may become relevant only if large scale
separations exists.

\subsection{RM-SOTON calculation}
\label{subsec:RM-SOTON}

The RM-SOTON collaboration has presented a strategy and preliminary results for
the calculation of the probability rate for the process
\begin{gather}
   h^- \to \ell^- + \bar{\nu}_\ell ( + \gamma )
\end{gather}
with $h=\pi,K$ at order
$\alpha_\text{em}$~\cite{Carrasco:2015xwa,Lubicz:2016xro,Lubicz:2016mpj,Tantalo:2016vxk}.
Notice that at this order, at most one photon can be produced in the final
state. In infinite volume, the relevant transition amplitudes have the following
behaviour in the $m_\gamma \to 0$ limit, which is obtained by expanding
eq.~\eqref{eq:IR:amplitude},
\begin{gather}
   M_{0\gamma}(m_\gamma) = \left( 1 + \frac{\alpha_\text{em}}{2} A \ln \frac{m_\gamma}{\mu} \right) \hat{M}_{0\gamma}^{(0)} + \alpha_\text{em} \hat{M}_{0\gamma}^{(1)}(\mu)
   + O(\alpha_\text{em} m_\gamma) + O(\alpha_\text{em}^2)
   \label{eq:RM:M0}
   \ , \\
   M_{1\gamma}(\vec{k};m_\gamma) = \alpha_\text{em}^{1/2} \hat{M}_{1\gamma}^{(0)}(\vec{k})
   + O(\alpha_\text{em}^{1/2} m_\gamma) + O(\alpha_\text{em}^{3/2})
   \ .
   \label{eq:RM:M1}
\end{gather}
As a special case of eq.~\eqref{eq:IR:transition-0}, the decay rate in the
center-of-mass frame and integrated over the soft photon is
\begin{gather}
   \Gamma(\Delta E,m_\gamma) = \Gamma_{0\gamma}(m_\gamma) + \Gamma_{1\gamma}(\Delta E,m_\gamma) + O(\alpha_\text{em}^2)
   \label{eq:RM:Gamma}
   \ , \\
   \Gamma_{0\gamma}(m_\gamma) = \frac{1}{2m_h} |M_{0\gamma}(m_\gamma)|^2
   \label{eq:RM:Gamma0}
   \ , \\
   \Gamma_{1\gamma}(\Delta E,m_\gamma)
   =
   \frac{1}{2m_h} \int_{|\vec{k}|<\Delta E} \frac{\d^3 k}{(2\pi)^3 2(m_\gamma^2+\vec{k}^2)^{1/2}} | M_{1\gamma}(\vec{k};m_\gamma) |^2
   \label{eq:RM:Gamma1}
   \ .
\end{gather}

The basic idea of the RM-SOTON collaboration is to exploit universality to
reshuffle the IR divergences in a convenient way. Let us consider the auxiliary
quantum field theory for hadron, leptons, neutrinos and photons, in which all
particles are described as fundamental fields and charged particles are
minimally-coupled to photons. In this auxiliary theory the electroweak
transition is generated at tree-level by an effective operator of dimension 5
(which can be found in~\cite{Tantalo:2016vxk}). This auxiliary quantum field
theory is usually referred to as \textit{point-like approximation}, as it could
be seen as the leading order of a general effective field theory. The
normalization of the effective EW Hamiltonian in the point-like approximation is
chosen in such a way that the tree-level amplitude matches the one in the
fundamental theory, i.e.
\begin{gather}
   \hat{M}_{\text{PT},0\gamma}^{(0)} = \hat{M}_{0\gamma}^{(0)} \ ,
   \label{eq:RM:matching}
\end{gather}
where the subscript PT denotes quantities calculated within the point-like
approximation. Then one defines the structure-dependent (SD) decay rates by
subtracting the point-like approximation to the full decay rate, i.e.
$\Gamma_{\text{SD},\star} = \Gamma_{\star} - \Gamma_{\text{PT},\star}$. In
particular
\begin{gather}
   \Gamma(\Delta E,m_\gamma)
   =
   \Gamma_\text{PT}(\Delta E,m_\gamma)
   +
   \Gamma_{\text{SD},0\gamma}(m_\gamma) + \Gamma_{\text{SD},1\gamma}(\Delta E,m_\gamma) + O(\alpha_\text{em}^2)
   \label{eq:RM:Gamma-deco}
   \ .
\end{gather}
The crucial observation is that the three terms in the r.h.s. of the above
equation are IR finite, and can be calculated separately.
\begin{enumerate}
   \item The first term in the r.h.s. of eq.~\eqref{eq:RM:Gamma-deco} is IR
   finite thanks to the Bloch-Nordsieck cancellation mechanism applied to the
   point-like approximation,
   \begin{gather}
      \Gamma_\text{PT}(\Delta E) \overset{\text{def}}{=}
      \lim_{m_\gamma \to 0} \Gamma_\text{PT}(\Delta E,m_\gamma) < \infty
      \ .
   \end{gather}
   Moreover this term is analytically calculable, given the tree-level
   amplitude, i.e. $F_h$.

   \item Let us look at second term in the r.h.s. of
   eq.~\eqref{eq:RM:Gamma-deco}. First we notice that a formula, similar to
   eq.~\eqref{eq:RM:M0}, holds for the transition amplitude calculated in the
   point-like approximation. Moreover the radiative function $A$ is the same in
   the full theory and in the point-like approximation, as it depends only on
   masses, charges and kinematics. Therefore
   \begin{gather}
      \Gamma_{\text{SD},0\gamma}
      \overset{\text{def}}{=}
      \lim_{m_\gamma \to 0} \Gamma_{\text{SD},0\gamma}(m_\gamma)
      =
      \frac{1}{2m_h} \lim_{m_\gamma \to 0}
      \left\{ |M_{0\gamma}(m_\gamma)|^2 - |M_{\text{PT},0\gamma}(m_\gamma)|^2 \right\}
      = \nonumber \\ \qquad =
      \frac{1}{2m_h} \lim_{m_\gamma \to 0} \left( 1 + \alpha_\text{em} \Re A \ln \frac{m_\gamma}{\mu} \right) \left\{ |\hat{M}_{0\gamma}^{(0)}|^2 - |\hat{M}_{\text{PT},0\gamma}^{(0)}|^2 \right\}
      + \nonumber \\ \qquad \qquad
      + \frac{\alpha_\text{em}}{m_h} \Re \left\{ [ \hat{M}_{0\gamma}^{(0)} ]^* \hat{M}_{0\gamma}^{(1)}(\mu)
      - [ \hat{M}_{\text{PT},0\gamma}^{(0)} ]^* \hat{M}_{\text{PT},0\gamma}^{(1)}(\mu) \right\}
      + O(\alpha_\text{em}^2)
      = \nonumber \\ \qquad =
      \frac{\alpha_\text{em}}{m_h} \Re [ \hat{M}_{0\gamma}^{(0)} ]^* [ \hat{M}_{0\gamma}^{(1)}(\mu) - \hat{M}_{\text{PT},0\gamma}^{(1)}(\mu) ]
      + O(\alpha_\text{em}^2) < \infty
      \ ,
   \end{gather}
   which shows that $\Gamma_{\text{SD},0\gamma}$ is IR finite as a consequence
   of the matching condition~\eqref{eq:RM:matching}. The RM-SOTON collaboration
   proposes to calculate $\Gamma_{\text{SD},0\gamma}$ from lattice simulations,
   where the box size $L$ acts as IR regulator. If we denote by
   $\tilde{\Gamma}_{\text{SD},0\gamma}(L)$ the decay rate in finite volume, the
   infinite-volume limit is found to be reached up to $1/L^2$
   corrections~\cite{Lubicz:2016xro}
   \begin{gather}
      \tilde{\Gamma}_{\text{SD},0\gamma}(L) = \Gamma_{\text{SD},0\gamma} + O(L^{-2}) \ .
      \label{eq:RM:L-2}
   \end{gather}
   This result is nontrivial, as both the point-like and full decay rate have a
   $1/L$ term (along with the $\ln L$ IR divergence) whose coefficient is fixed
   by the gauge Ward identities and general analyticity properties of the
   effective vertices.~\footnote{In fact this is a nontrivial consequence of the
   Low and Gell-Mann theorem~\cite{Low:1954kd,GellMann:1954kc}.}

   \item Finally the third term in the r.h.s. of eq.~\eqref{eq:RM:Gamma-deco} is
   IR finite since the l.h.s. of eq.~\eqref{eq:RM:Gamma-deco} is IR finite, i.e.
   \begin{gather}
      \Gamma_{\text{SD},1\gamma}(\Delta E) \overset{\text{def}}{=}
      \lim_{m_\gamma \to 0} \Gamma_{\text{SD},1\gamma}(\Delta E,m_\gamma) < \infty
      \ .
   \end{gather}
   If $\Delta E$ is asymptotically small, the PT contribution
   $\Gamma_\text{PT}(\Delta E)$ blows up logarithmically signalling break-down
   of the perturbative expansion, while the SD contribution
   $\Gamma_{\text{SD},1\gamma}(\Delta E)$ vanishes linearly. In the case of the
   pion decay, a window in $\Delta E$ exists such that both the
   $O(\alpha_\text{em}^2 \ln^2 \Delta E)$ corrections and
   $\Gamma_{\text{SD},1\gamma}(\Delta E)$ are negligible.
   It is shown in~\cite{Carrasco:2015xwa} that reasonable values of $\Delta E$
   for which this happens (independently of the lepton in the final state) are
   in the region of 10--50 MeV. The existence of such window is very much
   process dependent, and is expected not to hold in processes involving large
   scale separation because of large logarithms in the coefficients of the
   $\Delta E$ expansion. In general $\Gamma_{\text{SD},1\gamma}(\Delta E)$
   should be calculated from lattice simulations as well.

\end{enumerate}

The preliminary results presented by the RM-SOTON
collaboration~\cite{Lubicz:2016mpj} are obtained in the electroquenched setup
(i.e. photons couple only to the valence quarks, but not to the sea quarks),
with the QED$_\text{L}$ prescription. The infinite-volume extrapolation of the
SD contribution to the decay rate with no photons in the final state has been
performed by fitting the coefficient first unknown $L^{-2}$ correction,
according to the theoretical analysis presented in \cite{Lubicz:2016xro}.
Observables have been calculated with the RM123 method. For the $\pi \to \mu
+\nu$ process they report for instance
\begin{gather}
   \frac{\Gamma_{\text{NLO}}(\Delta E \simeq 30 \text{ MeV})}{\Gamma_{\text{LO}}} = 1.0210(15)(\dots)_\text{qQED} \ .
\end{gather}

\subsection{Large collinear logarithms}
\label{subsec:collinear}

The decomposition of decay rates in point-like and structure-dependent
contributions is completely general. At fixed order in the perturbative
expansion the point-like contribution contains the singular part in $\Delta E$,
which is the physical remnant of the soft divergences in the amplitudes. Soft
divergences are the only IR divergences in QED if all charged particles are
massive. In the limit in which some charged particles become massless new
divergences arise. As we will see, these divergences are generated by non-integrable angular
singularities in Feynman diagrams, and are therefore referred to as
\textit{collinear divergences}.

The simplest example of collinear divergences is found in the factor
$A_{\beta\alpha}$ in eq.~\eqref{eq:IR:Ah} for the process $h \to \ell + \nu$ in
the limit $m_\ell \to 0$, i.e.
\begin{gather}
   A_{\beta\alpha} \overset{m_\ell \ll m_h}{\simeq} \frac{2}{\pi} \ln \frac{m_h}{m_\ell} \to \infty \ .
   \label{eq:col:A}
\end{gather}
In the real world charged particles are massive. When the considered process
involves large scale separations large logarithms arise as a remnant of
collinear divergences. The divergence in eq.~\eqref{eq:col:A} arises from
virtual soft photons that are emitted in a direction parallel to the momentum of
the hard lepton. However collinear divergences arise also when the photon is not
soft, as we will try to argue with a specific example.

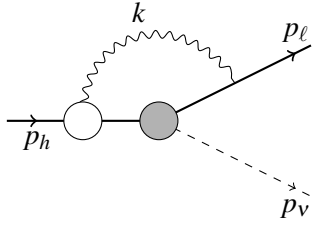
\begin{figure}
   
   \centering
   
   \tikzset{
     photon/.style={decorate, decoration={snake,amplitude=.5mm,segment length=1.8mm}, draw=black}
   }

   \begin{tikzpicture}
      
   \draw[thick,postaction={decorate},decoration={markings,mark=at position 0.2 with {\arrow{>}}}] (0,0) -- (2,0) node[pos=0.2,below] {$p_h$};
   \draw[thick,postaction={decorate},decoration={markings,mark=at position 0.9 with {\arrow{>}}}] (2,0) -- (4,1) node[pos=0.9,above] {$p_\ell$};
   \draw[dashed,postaction={decorate},decoration={markings,mark=at position 0.9 with {\arrow{>}}}] (2,0) -- (4,-1) node[pos=0.9,below] {$p_{\nu}$};
   \draw[fill=black!30] (2,0) circle (.25);
   \draw[photon] (1,0) .. controls (1,1.2) and (2.3,1.5) .. (3,.5) node[pos=0.5,above=1mm] {$k$};
   \draw[fill=white] (1,0) circle (.25);

   \end{tikzpicture}
   
   \label{fig:hdecay}
   \caption{Contribution with one virtual photon to the skeleton expansion of the $h^- \to \ell^- + \bar{\nu}_\ell$. The gray circle represents the insertion of the EW effective Hamiltonian, while the white circle represents the effective vertex $hh\gamma$.}

\end{figure}

Let us consider again the leptonic decay of the hadron $h^- \to \ell^- + \bar{\nu}_\ell$.
Let $\Gamma_\mu(p,k)$ be the effective vertex $hh\gamma$ where $k$ is the photon
incoming momentum, $p$ and $-p-k$ are the hadron incoming momenta, and let
\begin{gather}
   \Delta(p) = \frac{Z(p)}{p^2+m_h^2}
\end{gather}
be the dressed propagator of the hadron. Let us look at the particular
contribution to the transition probability represented in fig.~\ref{fig:hdecay}.
This contribution contains the integral
\begin{gather}
   N_\mu \int \frac{\d k_0}{2\pi} \int_{\Lambda_\text{IR} < |\vec{k}| < \Lambda_\text{UV}} \frac{\d^3 k}{(2\pi)^3} \frac{\Gamma_\mu(\bar{p}_h,k) \, Z(\bar{p}_h+k)}{k^2 (2 \bar{p}_h \cdot k + k^2) (2 \bar{p}_\ell \cdot k + k^2)}
   \ ,
\end{gather}
where Euclidean notation has been used, the on-shell four-momenta are
$\bar{p}_h=( i m_h,\vec{0} )$ and $\bar{p}_\ell=( i E_\ell,\vec{p}_\ell )$, and
$N_\mu$ is an overall normalization that depends on the kinematic variables,
polarizations and the Fermi constant. The IR cutoff has been introduced in order
to isolate the hard-photon contribution, and the UV cutoff has been introduced
in order to avoid confusion with UV divergences. The integral over $k_0$ can be
performed by closing the integration path at infinity in the complex plane. One
gets contributions from several poles. We are interested in the contribution
from either of the photon poles $\bar{k}=(\pm i |\vec{k}|, \vec{k})$ which is
\begin{gather}
   \frac {N_\mu }{8m_h} \int_{\Lambda_\text{IR} < |\vec{k}| < \Lambda_\text{UV}} \frac{\d^3 k}{(2\pi)^3} \frac{
      \Gamma_\mu(\bar{p}_h,\bar{k}) \, Z(\bar{p}_h+\bar{k})
   }{
      |\vec{k}|^2 \left( |\vec{k}| \sqrt{m_\ell^2 + \vec{p}_\ell^2} \mp \vec{p}_\ell \vec{k} \right)
   } \ .
\end{gather}
In spherical coordinates with $z = \cos \theta = \vec{p}_\ell \vec{k} / (
|\vec{p}_\ell| \, |\vec{k}| )$ it becomes apparent that the hadron propagator
has a non-integrable singularity at $z=\pm 1$ (the sign depends on which photon
pole we have chosen) for $m_\ell=0$, which translate into a logarithmic
divergence for $m_\ell$ small with respect to
$|\vec{p}_\ell|=(m_h^2-m_\ell^2)/(2m_h)$, or equivalently $m_\ell$ small with
respect to $m_h$,
\begin{gather}
   \frac {N_\mu }{8 (2\pi)^3 m_h |\vec{p}_\ell|} \int_{\Lambda_\text{IR}}^{\Lambda_\text{UV}} \frac{\d k}{k}
   \int_0^{2\pi} \d\phi \int_{-1}^1 \d z
   \frac{
      \Gamma_\mu(\bar{p}_h,\bar{k}) \, Z(\bar{p}_h+\bar{k})
   }{
      \sqrt{1 + \frac{m_\ell^2}{\vec{p}_\ell^2}} \mp z
   }
   \overset{m_\ell \ll m_h}{\simeq} \nonumber \\ \qquad \simeq
   \frac {N_\mu }{8 \pi^2 m_h^2} \int_{\Lambda_\text{IR}}^{\Lambda_\text{UV}} \frac{\d k}{k}
   \left[ \Gamma_\mu(\bar{p}_h,\bar{k}) \, Z(\bar{p}_h+\bar{k}) \right]_{\hat{\vec{k}}=\pm \hat{\vec{p}}_\ell}
   \ln \frac{m_h}{m_\ell}
   \ .
\end{gather}
This formula shows explicitly that the collinear divergence $\ln (m_h/m_\ell)$
gets contributions from virtual-photons poles with spatial momentum parallel to
the light-lepton momentum but with arbitrary modulus. In particular these
contributions are not universal, in the sense that they read the internal
structure of the hadron (encoded in the effective vertex and dressed propagator
away from the mass-shell).

The presence of quasi collinear divergences is only one of the mechanisms that
could enhance the structure-dependent part of the radiative corrections to decay
rates of heavy mesons. The existence of resonances slightly heavier than the
stable hadron, which can go almost on-shell in radiative
processes~\cite{Becirevic:2009aq,Bernlochner:2011bia} is another possible
mechanism. The inherently non-perturbative nature of these phenomena makes their
estimate quite difficult.

\section{Conclusions}

In order to be able to calculate hadronic observables at the level of the
percent precision, isospin-breaking corrections must be taken into account.
These have two sources: the up and down-quark mass difference and
electromagnetic interactions. The inclusion of QED effects in particular is
challenging for various reason.

Typically lattice simulations assume a finite box with periodic boundary
conditions along the spatial directions. However Gauss law forbids states with
total electric charge different from zero in a box with periodic boundary
conditions. This is an essential obstruction for numerical simulations which aim
at calculating properties of charged particles. I have reviewed some of the
proposed workarounds in section~\ref{sec:charged}. Among those, QED$_\text{SF}$
and QED$_\text{L}$ have been employed so far in large-scale simulations.
Unfortunately they are both non-local prescriptions. The non-locality in time
prevents a representation of the QED$_\text{SF}$ $n$-point functions in terms of
a transfer matrix. I have also argued with an explicit calculation in a
simplified model, that the QED$_\text{L}$ prescription destroys the
renormalization of composite operators (at least with dimension large than 4)
by local counterterms, and consequently the operator product expansion and the
Symanzik effective-theory description of the continuum limit. The theoretical
status of these non-local prescriptions is quite unsatisfactory.

My personal recommendation is to use and further develop theoretically sound
setups, as the only way to eliminate unwanted, uncontrollable and unnecessary
systematic errors in calculations which aim the the percent precision. The
QED$_\text{m}$ and QED$_\text{C}$ prescriptions are local, and can be studied
with the powerful machinery of QFT. In order to make QED$_\text{m}$ into a
practical tool, I personally feel that more work is necessary to understand the
implications of the non-commutativity of the $m_\gamma \to 0$ and $L \to \infty$
limits. My (surely biased) point of view is that our understanding of
QED$_\text{C}$ is mature enough to invest in a coding and numerical project.

The calculation of radiative corrections to decay rates (discussed in
section~\ref{sec:Smatrix}) presents extra complications with respect e.g. to
mass-splitting calculations due to the existence of IR divergences in QED.
Transition amplitudes beyond leading-order in $\alpha_\text{em}$ are IR
divergent because of loops of soft-photons. These divergences cancel in decay
rates, integrated over an arbitrary number of soft photons in the final state.
The separation of radiative corrections in a universal piece (a.k.a.
\textit{inner-bremsstrahlung}) and a \textit{structure-dependent} piece provides
the basis of a practical strategy for numerical calculations, put forward by the
RM-SOTON collaboration. An important fact in their analysis is that the
structure-dependent part of the real-photon emission decay rate can be neglected
in the leptonic decay of pion and kaon. In perspective one may be interested in
applying similar ideas to the decay of heavy mesons. On general grounds it is
possible to argue that structure-dependent contributions can be enhanced by the
presence of large logarithms due to quasi-collinear divergences or resonances
with a small off-shellness. In this case the real-photon emission is most
probably not negligible, and a detailed strategy to calculate this contribution
needs to be developed.

\acknowledgments

I wish to thank Nazario Tantalo and Alberto Ramos, colleagues and very good friends, for uncountable stimulating discussions about several topics touched in this contribution.

\appendix

\section{Effective mass in $\text{QED}_\text{TL}$}
\label{app:QEDTL}

In this appendix I want to derive eq.~\eqref{eq:QEDTL:divmeff}. In Fourier space
one gets the usual expression of the two-point function in terms of the
self-energy
\begin{gather}
   \tilde{C}(p_0) \overset{\text{def}}{=}
   \frac{1}{-i p_0 + m} - \frac{1}{-i p_0 + m} \Sigma_+(p_0) \frac{1}{-i p_0 + m} + O(e^4)
   \label{eq:QEDTL:Ctilde}
   \ , \\
   \Sigma_+(p_0)
   =
   \frac{e^2}{TL^3} \sum_{\substack{k \in \frac{2\pi}{T} \mathbb{Z} \times \frac{2\pi}{L} \mathbb{Z}^3 \\ k\neq 0}} \frac{-i(p_0+k_0) + 2m}{(p_0+k_0)^2 + \vec{k}^2 + m^2} \frac{\Lambda^2}{k^2(k^2+\Lambda^2)}
   \ ,
   \label{eq:QEDTL:Sigma}
\end{gather}
where a Pauli-Villard regulator has been introduced for the photon propagator.
The exclusion of the zero momentum in the sum of the self-energy is a direct
consequence of the constraint defining $\text{QED}_\text{TL}$. Notice that, even
though eqs.~\eqref{eq:QEDTL:Ctilde} and~\eqref{eq:QEDTL:Sigma} define
$\tilde{C}(p_0)$ and $\Sigma_+(p_0)$ for all real values of $p_0$, the simple
relation
\begin{gather}
   \tilde{C}(p_0) = \int_0^T \d t \, e^{ip_0 t} \, C(t)
   \label{eq:QEDTL:Ctilde-1}
\end{gather}
is valid only for momenta $p_0$ allowed by the boundary conditions, i.e.
\begin{gather}
   p_0 \in \frac{2\pi}{T} \mathbb{Z} \ .
\end{gather}

Before transforming back to the time-momentum representation, let us study the
pole structure of $\tilde{C}(p_0)$. The self-energy is analytical in a complex
neighbourhood of $p_0=-im$, and can be written as
\begin{gather}
   \Sigma_+(p_0) = \Sigma_+(-im) + (-i p_0 + m) i \Sigma_+'(-im) - (-i p_0 + m)^2 R(p_0) \ .
\end{gather}
Explicit calculation shows that the reminder $R(p_0)$, as the self-energy
itself, can be represented as an infinite sum of single poles away from
$p_0=-im$,
\begin{gather}
   R(p_0) = \sum_N \frac{r_N}{-i p_0 + z_N} \ .
\end{gather}
One can plug this representation for the self-energy into
eq.~\eqref{eq:QEDTL:Ctilde}, and then back into eq.~\eqref{eq:QEDTL:C}, use the
Poisson summation formula, and obtain the two-point function in time-momentum
representation by explicit calculation
\begin{gather}
   C(t) = \frac{1}{T} \sum_{p_0 \in \frac{2\pi}{T} \mathbb{Z}} e^{-ip_0 t} \, \tilde{C}(p_0)
   =
   \sum_{n \in \mathbb{Z}} \int \frac{\d p_0}{2\pi} e^{-ip_0 (t+nT)} \, \tilde{C}(p_0)
   = \nonumber \\ \qquad =
   \frac{e^{-tm}}{1- e^{-Tm}} \left\{ 1 -  i \Sigma_+'(-im) - \left( t+ \frac{T}{e^{Tm}-1} \right) \Sigma_+(-im) \right\}
   + \sum_N \frac{ r_N e^{-t z_N} }{1-e^{-T z_N}}
   + O(e^4)
   \label{eq:QEDTL:C-1}
   \ .
\end{gather}
This representation is valid for $0<t<T$. The effective mass is readily
calculated
\begin{gather}
   m_\text{eff}(t) = - \frac{\d}{\d t} \ln C(t) = m + \Sigma_+(-im)
   + \sum_N \frac{r_N (1- e^{-Tm})}{1-e^{-T z_N}} (z_N-m) e^{-t (z_N-m)}
   + O(e^4) \ .
   \label{eq:QEDTL:meff-1}
\end{gather}
Notice that the on-shell self-energy $\Sigma_+(-im)$ as well as the residues
$r_N$ and the poles $z_N$ depend on $T$. We will show that the
effective mass is divergent in the $T \to \infty$ limit, by showing that the
on-shell self-energy is.

Some tedious manipulation gives the following decomposition for the on-shell
self-energy
\begin{gather}
   \Sigma_+(-im)
   =
   \frac{m e^2}{TL^3} \sum_{\substack{k \in \frac{2\pi}{T} \mathbb{Z} \times \frac{2\pi}{L} \mathbb{Z}^3 \\ k\neq 0}} \frac{2k_0^2+k^2}{4m^2k_0^2+k^4} \frac{\Lambda^2}{k^2(k^2+\Lambda^2)}
   = \nonumber \\ \qquad =
   \frac{3 e^2}{4 m TL^3} \sum_{k_0 \in \frac{2\pi}{T} \mathbb{Z}/\{0\}} \frac{1}{k_0^2}
   - \frac{3 m e^2}{TL^3} \sum_{k_0 \in \frac{2\pi}{T} \mathbb{Z}/\{0\}} \frac{1}{4m^2 + k_0^2} \left( \frac{1}{k_0^2+\Lambda^2} + \frac{1}{4m^2} \right)
   + \nonumber \\ \qquad \qquad
   + \frac{m e^2}{TL^3} \sum_{k_0 \in \frac{2\pi}{T} \mathbb{Z}} \sum_{\vec{k} \in \frac{2\pi}{L} \mathbb{Z}^3 / \{\vec{0}\} } \frac{2k_0^2+k^2}{4m^2k_0^2+k^4} \frac{\Lambda^2}{k^2(k^2+\Lambda^2)}
   \ .
   \label{eq:QEDTL:Sigma-onshell}
\end{gather}
In the $T \to \infty$, the first term after the last equality is divergent
(while the other terms are finite).
\begin{gather}
   \Sigma_+(-im)
   =
   \frac{3 e^2 T}{8 \pi^2 m L^3} \sum_{n=1}^\infty \frac{1}{n^2} +  O(T^0)
   =
   \frac{e^2 T}{16 m L^3} + O(T^0) \ .
   \label{eq:QEDTL:Sigma-T}
\end{gather}
This shows that the effective mass given in eq.~\eqref{eq:QEDTL:meff-1} is
divergent in the $T \to \infty$.

\section{Harmonic oscillator with SF boundary conditions}
\label{app:harmo}

The harmonic oscillator Hamiltonian
\begin{gather}
   H = \frac{p^2}{2} + \frac{\lambda}{2} q^2
   \label{eq:QEDTL:HAO}
\end{gather}
has a discrete spectrum for any $\lambda >0$. If $|\text{SF}\rangle$ is the
eigenstate of the position operator $q$ corresponding to zero eigenvalue, one
can define the correlator with Schr\"odinger-functional boundary conditions, and
the related effective mass as
\begin{gather}
   C^\text{AO}(t) \overset{\text{def}}{=} \langle \text{SF} | e^{-(\frac{T}{2}-t) H} q e^{-tH} q e^{- \frac{T}{2} H} | \text{SF} \rangle
   \label{eq:QEDTL:CAO}
   \ , \\
   m^\text{AO}_\text{eff}(t) \overset{\text{def}}{=} - \frac{\d}{\d t} \ln C^\text{AO}(t)
   =
   \frac{
   \langle \text{SF} | e^{-(\frac{T}{2}-t) H} [q,H] e^{-tH} q e^{- \frac{T}{2} H} | \text{SF} \rangle
   }{
   \langle \text{SF}| e^{-(\frac{T}{2}-t) H} q e^{-tH} q e^{- \frac{T}{2} H} | \text{SF} \rangle
   }
   \ .
\end{gather}
In the $T \to \infty$ limit, the evolution operator $e^{- \frac{T}{2} H}$
projects over the (unperturbed) vacuum $| \Omega \rangle$, i.e.
\begin{gather}
   \lim_{T \to \infty} m^\text{AO}_\text{eff}(t)
   =
   \frac{
   \langle \Omega | q (H-E_0) e^{-tH} q |\Omega \rangle 
   }{
   \langle \Omega | q e^{-tH} q  | \Omega \rangle
   }
   \ ,
\end{gather}
which is clearly finite. In the $t \to \infty$ limit, as expected, the
effective mass at $T=\infty$ approaches $E_1-E_0$, where $E_0$ and $E_1$ are the
energies of vacuum and first excited state respectively.

Notice that the Hamiltonian in eq.~\eqref{eq:QEDTL:HAO} does not have a regular
perturbative expansion in $\lambda$ (in particular the eigenvalues are
proportional to $\lambda^{1/2}$). On the other hand the correlator can be
represented in terms of a path integral with Schr\"odinger functional boundary
conditions
\begin{gather}
   C^\text{AO}(t) = \frac{1}{Z} \int_{q(-T/2)=q(T/2)=0} [ \d q ] \ e^{-S} q(t) q(0) \ ,
\end{gather}
where the action is simply given by
\begin{gather}
   S = \int_{-T/2}^{T/2} \d t \ \left\{ \frac{\dot{q}^2}{2} + \frac{\lambda}{2} q^2 \right\} \ .
   \label{eq:QEDTL:SAO}
\end{gather}
The boundary conditions kill the constant mode of $q(t)$ in a local fashion, and
make the action with $\lambda=0$ strictly positive. The correlator and the
effective mass have a well-defined perturbative expansion as long as $T$ is
finite, and can be calculated with standard Feynman diagram techniques.

The quadratic part of the action~\eqref{eq:QEDTL:SAO} defines the free
propagator
\begin{gather}
   \Delta(t,s)
   =
   \frac{2}{T} \sum_{p \in \frac{\pi}{T} (\mathbb{N}+1) }\frac{1}{p^2} \sin \left[ p \left( t+\tfrac{T}{2} \right) \right] \ \sin \left[ p \left( s+\tfrac{T}{2} \right) \right]
   \ .
\end{gather}
Some algebraic manipulation yields the following representation
\begin{gather}
   \Delta(t,s)
   =
   \frac{T}{4}
   -\frac{1}{T} \sum_{p \in \frac{\pi}{T} \mathbb{Z} / \{0\} } \frac{1}{p^2} \sin^2 \frac{p (t-s)}{2}
   + \nonumber \\ \qquad \qquad
   +\frac{1}{T} \left( \sum_{p \in \frac{2\pi}{T} \mathbb{Z} / \{0\} } - \sum_{p \in \frac{\pi}{T} (2\mathbb{Z}+1) } \right) \frac{1}{p^2} \sin^2 \frac{p (s+t)}{2}
   \ ,
\end{gather}
which is more convenient to take the $T \to \infty$ limit,
\begin{gather}
   \Delta(t,s)
   \overset{T \to \infty}{=}
   \frac{T}{4}
   - \frac{1}{\pi} \int_{-\infty}^\infty \frac{\d p}{p^2} \sin^2 \frac{p (t-s)}{2}
   - \frac{ts}{T}
   + {\dots}
   =
   \frac{T}{4} - \frac{1}{2} |t-s| - \frac{ts}{T}
   + {\dots} \ ,
\end{gather}
where the dots stand for corrections that fall off faster than any inverse power
of $T$. Notice that the free propagator contains an IR divergence and it is not
defined in the $T \to \infty$ limit, as a consequence of the accumulation of the
eigenvalues of $-\d^2/\d t^2$ in zero. However this divergence drops out the
effective mass.

At first order in $\lambda$, the two-point function defined in
eq.~\eqref{eq:QEDTL:CAO} is
\begin{gather}
   C^\text{AO}(t) = \Delta(t,0) - \lambda \int_{-T/2}^{T/2} \d s \ \Delta(t,s) \Delta(0,s) + O(\lambda^2)
   = \nonumber \\ \qquad =
   \frac{T}{4} - \frac{1}{2} |t| 
   - \lambda \left\{
   \frac{T^3}{48} - \frac{T t^2}{8} + \frac{|t|^3}{12}
   \right\}
   + O(\lambda^2) + \dots
   \ ,
\end{gather}
and the effective mass
\begin{flalign}
   m^\text{AO}_\text{eff}(t) = &
   - \frac{\partial_t \Delta(t,0)}{\Delta(t,0)}
   - \frac{\lambda \partial_t \Delta(t,0)}{[\Delta(t,0)]^2} \int_{-T/2}^{T/2} \d s \ \Delta(t,s) \Delta(0,s)
   + \nonumber \\ & \qquad
   + \frac{\lambda}{\Delta(t,0)} \int_{-T/2}^{T/2} \d s \ \partial_t \Delta(t,s) \Delta(0,s) + O(\lambda^2)
   = \nonumber \\ = &
   \frac{\lambda T}{6} \text{sgn}(t) - \frac{\lambda}{3} t + O(T^{-1}) + O(\lambda^2)
   \ .
\end{flalign}
Notice that the effective mass at $O(\lambda)$ is divergent in the $T \to
\infty$ limit, even if the non-perturbative two-point function has a Hamiltonian
representation.

\section{Fermion two-point function in $\text{QED}_\text{m}$}
\label{app:QEDm}

In this appendix we will be concerned only with the fermion two-point function,
even though most results extend also to more complicated observables.

Instead of considering the full path integral, we consider only the
fermion-field integration. Large gauge transformations~\eqref{eq:gen:large-def}
act as a spurionic symmetry on the fermionic path integral, i.e. the following
equality is valid for any $n \in \mathbb{Z}^4$
\begin{gather}
   \label{eq:QEDm:fermionic1}
   \int [\d \psi] \, [\d \bar{\psi}] \  \psi(x) \bar{\psi}(0)
   e^{- \int \d^4 z \ \bar{\psi} (\gamma_\mu D_\mu[\alpha,B] + m) \psi(z)}
   = \\ \qquad =
   e^{2\pi i (L^{-1}n)_\mu x_\mu}
   \int [\d \psi] \, [\d \bar{\psi}] \  \psi(x) \bar{\psi}(0)
   e^{- \int \d^4 z \ \bar{\psi} ( \gamma_\mu D_\mu[\alpha + 2\pi n, B] + m ) \psi(z)}
   \ , \nonumber
\end{gather}
where I have separated the dependence on constant mode and fluctuation in the
covariant derivative $D_\mu[\alpha,B]$. Let us consider the change of variables
\begin{gather}
   \psi(x) \to e^{-i (\alpha_\mu+ 2\pi n_\mu) (L^{-1}x)_\mu} \psi(x) \ , \nonumber \\
   \bar{\psi}(x) \to e^{i (\alpha_\mu+ 2\pi n_\mu) (L^{-1}x)_\mu} \bar{\psi}(x) \ ,
   \label{eq:QEDm:cov}
\end{gather}
defined in $0 < x_\mu < L_\mu$ and extended by periodicity elsewhere. This means
that the fermionic boundary conditions do not change, but the fermionic action
gets a singular contribution. In particular
\footnote{
The discontinuous exponential $e^{-i (\alpha_\mu+ 2\pi n_\mu) (L^{-1}x)_\mu}$
must be defined through a proper limiting procedure. We consider a family of
functions $f_\epsilon(\hat{x})$. For each $\epsilon >0$ we require that
$f_\epsilon(\hat{x})$ is smooth and periodic. We also require that
\begin{gather}
   \lim_{\epsilon \to 0^+} f_\epsilon(\hat{x}) = \sum_{k \in \mathbb{Z}} (\hat{x}-k) \chi_{(k,k+1)}(\hat{x}) \ , \\
   \lim_{\epsilon \to 0^+} f'_\epsilon(\hat{x}) = 1 - \sum_{k \in \mathbb{Z}} \delta(\hat{x}-k) \ ,
\end{gather}
where $\chi_A(x)$ is the characteristic function of the set $A$. The new
fermionic fields are defined as
\begin{gather}
   \eta_\epsilon(x) = e^{i \sum_\mu (\alpha_\mu+ 2\pi n_\mu) f_\epsilon(\tfrac{x_\mu}{L_\mu})} \psi(x) \ , \quad
   \bar{\eta}_\epsilon(x) = e^{-i \sum_\mu (\alpha_\mu+ 2\pi n_\mu) f_\epsilon(\tfrac{x_\mu}{L_\mu})} \bar{\psi}(x_\mu/L_\mu) \ ,
\end{gather}
and they satisfy periodic boundary conditions. In terms of the new fields the
action in the r.h.s. of eq.~\eqref{eq:QEDm:fermionic1} becomes
\begin{gather}
   \bar{\psi} ( \gamma_\mu D_\mu[\alpha + 2\pi n, B] + m ) \psi(z)
   = \nonumber \\ \qquad =
   \bar{\eta}_\epsilon ( \gamma_\mu D_\mu[0, B] + m ) \eta_\epsilon(z)
   + i \sum_\mu L^{-1}_\mu (\alpha_\mu + 2\pi n_\mu) [ 1 - f'_\epsilon(\tfrac{x_\mu}{L_\mu}) ] \bar{\eta}_\epsilon \gamma_\mu \eta_\epsilon(z)
   = \nonumber \\ \quad \overset{\epsilon \to 0^+} =
   \bar{\eta}_0 ( \gamma_\mu D_\mu[0, B] + m ) \eta_0(z)
   + i \sum_{k \in \mathbb{Z}} \sum_\mu (\alpha_\mu + 2\pi n_\mu) \delta(x_\mu-k L_\mu) \bar{\eta}_0 \gamma_\mu \eta_0(z)
\end{gather}
Eq.~\eqref{eq:QEDm:ferm-action} is obtained by integrating over $z$ and by
replacing $\eta_0 \to \psi$ and $\bar{\eta}_0 \to \bar{\psi}$.
}
\begin{gather}
   \label{eq:QEDm:ferm-action}
   \int \d^4 z \ \bar{\psi} ( \gamma_\mu D_\mu[\alpha + 2\pi n, B] + m ) \psi(z)
   \to \\ \qquad \to
   \int \d^4 z \ \bar{\psi} ( \gamma_\mu D_\mu[0, B] + m ) \psi(z)
   + i (\alpha_\mu + 2\pi n_\mu) Q_\mu
   \ , \nonumber
\end{gather}
where $Q_\mu$ is the electric charge (in units of $e$) operator defined by
interpreting $\mu$ as temporal direction and localized on the time-slice
$x_\mu=0$, i.e.
\begin{gather}
   Q_\mu = \int \d^4 z \ \delta(z_\mu) \bar{\psi} \gamma_\mu \psi(z) \ .
\end{gather}
Therefore, by using the change of variables~\eqref{eq:QEDm:cov} into the r.h.s.
of eq.~\eqref{eq:QEDm:fermionic1} one obtains
\begin{gather}
   \label{eq:QEDm:fermionic2}
   \int [\d \psi] \, [\d \bar{\psi}] \  \psi(x) \bar{\psi}(0^+)
   e^{- \int \d^4 z \ \bar{\psi} (\gamma_\mu D_\mu[\alpha,B] + m) \psi(z)}
   = \\ \qquad =
   \int [\d \psi] \, [\d \bar{\psi}] \  \psi(x) \bar{\psi}(0^+)
   e^{- \int \d^4 z \ \bar{\psi} ( \gamma_\mu D_\mu[0, B] + m ) \psi(z)}
   e^{- i \alpha_\mu (Q+L^{-1}x)_\mu} e^{ -  2\pi i n_\mu Q_\mu}
   \ . \nonumber
\end{gather}
Since this equality has to be valid for any $n \in \mathbb{Z}^4$, it means that
$Q_\mu$ must have support on integer numbers \textit{after} the fermionic path
integral is evaluated, therefore one can write
\begin{gather}
   \label{eq:QEDm:fermionic3}
   \int [\d \psi] \, [\d \bar{\psi}] \  \psi(x) \bar{\psi}(0^+)
   e^{- \int \d^4 z \ \bar{\psi} (\gamma_\mu D_\mu[\alpha,B] + m) \psi(z)}
   = \\ \qquad =
   \sum_{q \in \mathbb{Z}^4} e^{- i \alpha_\mu (q+L^{-1}x)_\mu}  \int [\d \psi] \, [\d \bar{\psi}] \  \psi(x) \bar{\psi}(0^+)
   e^{- \int \d^4 z \ \bar{\psi} ( \gamma_\mu D_\mu[0, B] + m ) \psi(z)}
   \delta_{Q,q}
   \ , \nonumber
\end{gather}
where the Kr\"onecker delta should be interpreted in the distributional sense
\begin{gather}
   \delta_{Q,q} = \int_{(-\epsilon,\epsilon)^4} \d^4 z \ \delta^4(Q-q-z)
   =
   \int_{(-\epsilon,\epsilon)^4} \d^4 z \int \frac{\d^4 \xi}{(2\pi)^4} \ e^{i \xi(Q-q-z)}
   \ .
\end{gather}

Let us take one step back, and comment the result. This long derivation has
yielded a very natural result: the electric charge takes integer values. This
fact is obvious if we consider the lattice-discretized theory in Hamiltonian
formalism. However in the path-integral formalism it is a nontrivial consequence
of large-gauge transformations.

We can use the representation~\eqref{eq:QEDm:fermionic3} in the full
QED$_\text{m}$ path-integral with action~\eqref{eq:QEDm:action}, together with
the decomposition given in eq.~\eqref{eq:gen:decomposition}, and obtain
\begin{gather}
   \int [\d A] \, [\d \psi] \, [\d \bar{\psi}] \  \psi(x) \bar{\psi}(0^+)
   e^{- S_\text{m}(A,\psi,\bar{\psi}) }
   = 
   \sum_{q \in \mathbb{Z}^4} 
   \int \d^4 \alpha \ e^{-\frac{m_\gamma^2 V}{2 e^2} (L^{-1}\alpha)_\mu^2} e^{- i \alpha_\mu (q+L^{-1}x)_\mu}
   \times \nonumber \\ \qquad \qquad \times
   \int [\d B] \, [\d \psi] \, [\d \bar{\psi}] \  \psi(x) \bar{\psi}(0^+)
   e^{- S(B,\psi,\bar{\psi}) - \frac{m_\gamma^2}{2} \int \d^4 z \ B^2_\mu(z)} \delta_{Q,q}
   \ .
\end{gather}
At this point we notice that: \textit{(1)} the integral over the constant mode
is Gaussian and can be explicitly calculated; \textit{(2)} the remaining
integral over fluctuations and fermions is nothing but a path-integral in
QED$_\text{TL}$, i.e.
\begin{gather}
   \int [\d A] \, [\d \psi] \, [\d \bar{\psi}] \  \psi(x) \bar{\psi}(0^+)
   e^{- S_\text{m}(A,\psi,\bar{\psi}) }
   = \nonumber \\ \qquad =
   \frac{(2\pi)^2 e^4 Z_{\text{TL}}}{m_\gamma^4 V} \sum_{q \in \mathbb{Z}^4} 
   e^{- \frac{e^2}{2 m_\gamma^2 V} (Lq+x)^2 }
   \langle \psi(x) \bar{\psi}(0^+)
   e^{- \frac{m_\gamma^2}{2} \int \d^4 z \ B^2_\mu(z)} \delta_{Q,q}
   \rangle_\text{TL}
   \ .
\end{gather}
A similar representation can be proven to hold for
the partition function, which yields for the two-point function in QED$_\text{m}$
\begin{gather}
   \langle \psi(x) \bar{\psi}(0^+) \rangle_\text{m} 
   =
   \frac{
   \sum_{q \in \mathbb{Z}^4} 
   e^{- \frac{e^2}{2 m_\gamma^2 V} (Lq+x)^2 }
   \langle \psi(x) \bar{\psi}(0^+)
   e^{- \frac{m_\gamma^2}{2} \int \d^4 z \ B^2_\mu(z)} \delta_{Q,q}
   \rangle_\text{TL}
   }{
   \sum_{q \in \mathbb{Z}^4} 
   e^{- \frac{e^2}{2 m_\gamma^2 V} (Lq)^2 }
   \langle e^{- \frac{m_\gamma^2}{2} \int \d^4 z \ B^2_\mu(z)} \delta_{Q,q}
   \rangle_\text{TL}
   }
   \ .
\end{gather}

\bibliographystyle{JHEP}
\bibliography{patella}

\providecommand{\href}[2]{#2}\begingroup\raggedright\begin{thebibliography}{10}

\bibitem{Duncan:1996xy}
A.~Duncan, E.~Eichten and H.~Thacker, \emph{{Electromagnetic splittings and
  light quark masses in lattice QCD}},
  \href{http://dx.doi.org/10.1103/PhysRevLett.76.3894}{\emph{Phys. Rev. Lett.}
  {\bf 76} (1996) 3894--3897},
  [\href{https://arxiv.org/abs/hep-lat/9602005}{{\tt hep-lat/9602005}}].

\bibitem{Basak:2013iw}
{\scshape MILC} collaboration, S.~Basak et~al., \emph{{Electromagnetic
  contributions to pseudoscalar masses}}, {\emph{PoS} {\bf CD12} (2013) 030},
  [\href{https://arxiv.org/abs/1301.7137}{{\tt 1301.7137}}].

\bibitem{Blum:2010ym}
T.~Blum, R.~Zhou, T.~Doi, M.~Hayakawa, T.~Izubuchi, S.~Uno et~al.,
  \emph{{Electromagnetic mass splittings of the low lying hadrons and quark
  masses from 2+1 flavor lattice QCD+QED}},
  \href{http://dx.doi.org/10.1103/PhysRevD.82.094508}{\emph{Phys. Rev.} {\bf
  D82} (2010) 094508}, [\href{https://arxiv.org/abs/1006.1311}{{\tt
  1006.1311}}].

\bibitem{Portelli:2012pn}
A.~Portelli et~al., \emph{{Systematic errors in partially-quenched QCD plus QED
  lattice simulations}}, {\emph{PoS} {\bf LATTICE2011} (2011) 136},
  [\href{https://arxiv.org/abs/1201.2787}{{\tt 1201.2787}}].

\bibitem{Borsanyi:2014jba}
S.~Borsanyi et~al., \emph{{Ab initio calculation of the neutron-proton mass
  difference}}, \href{http://dx.doi.org/10.1126/science.1257050}{\emph{Science}
  {\bf 347} (2015) 1452--1455}, [\href{https://arxiv.org/abs/1406.4088}{{\tt
  1406.4088}}].

\bibitem{Horsley:2015vla}
R.~Horsley et~al., \emph{{QED effects in the pseudoscalar meson sector}},
  \href{http://dx.doi.org/10.1007/JHEP04(2016)093}{\emph{JHEP} {\bf 04} (2016)
  093}, [\href{https://arxiv.org/abs/1509.00799}{{\tt 1509.00799}}].

\bibitem{Horsley:2015eaa}
R.~Horsley et~al., \emph{{Isospin splittings of meson and baryon masses from
  three-flavor lattice QCD + QED}},
  \href{http://dx.doi.org/10.1088/0954-3899/43/10/10LT02}{\emph{J. Phys.} {\bf
  G43} (2016) 10LT02}, [\href{https://arxiv.org/abs/1508.06401}{{\tt
  1508.06401}}].

\bibitem{Young:LAT2016}
R.~Young, \emph{{Infrared features of dynamical QED+QCD simulations}},  in
  \emph{{Proceedings, 34th International Symposium on Lattice Field Theory
  (Lattice 2016): Southampton, UK, July 24-30, 2016}}, 2016.

\bibitem{Liu:2016kbb}
C.~Liu, \emph{{Review on Hadron Spectroscopy}},  in \emph{{Proceedings, 34th
  International Symposium on Lattice Field Theory (Lattice 2016): Southampton,
  UK, July 24-30, 2016}}, 2016.
\newblock \href{https://arxiv.org/abs/1612.00103}{{\tt 1612.00103}}.

\bibitem{Fodor:2016bgu}
Z.~Fodor, C.~Hoelbling, S.~Krieg, L.~Lellouch, T.~Lippert, A.~Portelli et~al.,
  \emph{{Up and down quark masses and corrections to Dashen's theorem from
  lattice QCD and quenched QED}},
  \href{http://dx.doi.org/10.1103/PhysRevLett.117.082001}{\emph{Phys. Rev.
  Lett.} {\bf 117} (2016) 082001},
  [\href{https://arxiv.org/abs/1604.07112}{{\tt 1604.07112}}].

\bibitem{Varnhorst:LAT2016}
L.~Varnhorst, \emph{{Up and down quark masses and corrections to Dashen's
  theorem from lattice QCD and quenched QED}},  in \emph{{Proceedings, 34th
  International Symposium on Lattice Field Theory (Lattice 2016): Southampton,
  UK, July 24-30, 2016}}, 2016.

\bibitem{Aoki:2016frl}
S.~Aoki et~al., \emph{{Review of lattice results concerning low-energy particle
  physics}},  \href{https://arxiv.org/abs/1607.00299}{{\tt 1607.00299}}.

\bibitem{Rosner:2015wva}
J.~L. Rosner, S.~Stone and R.~S. Van~de Water, \emph{{Leptonic Decays of
  Charged Pseudoscalar Mesons - 2015}}, {\emph{Submitted to: Particle Data
  Book} (2015) }, [\href{https://arxiv.org/abs/1509.02220}{{\tt 1509.02220}}].

\bibitem{Cirigliano:2011ny}
V.~Cirigliano, G.~Ecker, H.~Neufeld, A.~Pich and J.~Portoles, \emph{{Kaon
  Decays in the Standard Model}},
  \href{http://dx.doi.org/10.1103/RevModPhys.84.399}{\emph{Rev. Mod. Phys.}
  {\bf 84} (2012) 399}, [\href{https://arxiv.org/abs/1107.6001}{{\tt
  1107.6001}}].

\bibitem{Carrasco:2015xwa}
N.~Carrasco, V.~Lubicz, G.~Martinelli, C.~T. Sachrajda, N.~Tantalo,
  C.~Tarantino et~al., \emph{{QED Corrections to Hadronic Processes in Lattice
  QCD}}, \href{http://dx.doi.org/10.1103/PhysRevD.91.074506}{\emph{Phys. Rev.}
  {\bf D91} (2015) 074506}, [\href{https://arxiv.org/abs/1502.00257}{{\tt
  1502.00257}}].

\bibitem{Lubicz:2016xro}
V.~Lubicz, G.~Martinelli, C.~T. Sachrajda, F.~Sanfilippo, S.~Simula and
  N.~Tantalo, \emph{{Finite-Volume QED Corrections to Decay Amplitudes in
  Lattice QCD}},  \href{https://arxiv.org/abs/1611.08497}{{\tt 1611.08497}}.

\bibitem{Lubicz:2016mpj}
V.~Lubicz, G.~Martinelli, C.~T. Sachrajda, F.~Sanfilippo, S.~Simula, N.~Tantalo
  et~al., \emph{{Electromagnetic corrections to the leptonic decay rates of
  charged pseudoscalar mesons: lattice results}},  in \emph{{Proceedings, 34th
  International Symposium on Lattice Field Theory (Lattice 2016): Southampton,
  UK, July 24-30, 2016}}, 2016.
\newblock \href{https://arxiv.org/abs/1610.09668}{{\tt 1610.09668}}.

\bibitem{Tantalo:2016vxk}
N.~Tantalo, V.~Lubicz, G.~Martinelli, C.~T. Sachrajda, F.~Sanfilippo and
  S.~Simula, \emph{{Electromagnetic corrections to leptonic decay rates of
  charged pseudoscalar mesons: finite-volume effects}},
  \href{https://arxiv.org/abs/1612.00199}{{\tt 1612.00199}}.

\bibitem{Boyle:2016lbc}
P.~Boyle, V.~Gülpers, J.~Harrison, A.~Jüttner, A.~Portelli and C.~Sachrajda,
  \emph{{Electromagnetic Corrections to Meson Masses and the HVP}},  in
  \emph{{Proceedings, 34th International Symposium on Lattice Field Theory
  (Lattice 2016): Southampton, UK, July 24-30, 2016}}, 2016.
\newblock \href{https://arxiv.org/abs/1612.05962}{{\tt 1612.05962}}.

\bibitem{Olive:2016xmw}
{\scshape Particle Data Group} collaboration, C.~Patrignani et~al.,
  \emph{{Review of Particle Physics}},
  \href{http://dx.doi.org/10.1088/1674-1137/40/10/100001}{\emph{Chin. Phys.}
  {\bf C40} (2016) 100001}.

\bibitem{Bailey:2014tva}
{\scshape Fermilab Lattice, MILC} collaboration, J.~A. Bailey et~al.,
  \emph{{Update of $|V_{cb}|$ from the $\bar{B}\to D^*\ell\bar{\nu}$ form
  factor at zero recoil with three-flavor lattice QCD}},
  \href{http://dx.doi.org/10.1103/PhysRevD.89.114504}{\emph{Phys. Rev.} {\bf
  D89} (2014) 114504}, [\href{https://arxiv.org/abs/1403.0635}{{\tt
  1403.0635}}].

\bibitem{Aubert:2007qs}
{\scshape BaBar} collaboration, B.~Aubert et~al., \emph{{Measurement of the
  Decay $B^{-} \to$ D*0 $e^{-} \bar{\nu}$( $e$)}},
  \href{http://dx.doi.org/10.1103/PhysRevLett.100.231803}{\emph{Phys. Rev.
  Lett.} {\bf 100} (2008) 231803}, [\href{https://arxiv.org/abs/0712.3493}{{\tt
  0712.3493}}].

\bibitem{Adam:2002uw}
{\scshape CLEO} collaboration, N.~E. Adam et~al., \emph{{Determination of the
  anti-B ---> D* l anti-nu decay width and |V(cb)|}},
  \href{http://dx.doi.org/10.1103/PhysRevD.67.032001}{\emph{Phys. Rev.} {\bf
  D67} (2003) 032001}, [\href{https://arxiv.org/abs/hep-ex/0210040}{{\tt
  hep-ex/0210040}}].

\bibitem{Aubert:2008yv}
{\scshape BaBar} collaboration, B.~Aubert et~al., \emph{{Measurements of the
  Semileptonic Decays anti-B ---> D l anti-nu and anti-B ---> D* l anti-nu
  Using a Global Fit to D X l anti-nu Final States}},
  \href{http://dx.doi.org/10.1103/PhysRevD.79.012002}{\emph{Phys. Rev.} {\bf
  D79} (2009) 012002}, [\href{https://arxiv.org/abs/0809.0828}{{\tt
  0809.0828}}].

\bibitem{Amhis:2012bh}
{\scshape Heavy Flavor Averaging Group} collaboration, Y.~Amhis et~al.,
  \emph{{Averages of B-Hadron, C-Hadron, and tau-lepton properties as of early
  2012}},  \href{https://arxiv.org/abs/1207.1158}{{\tt 1207.1158}}.

\bibitem{Bernlochner:2011bia}
F.~Bernlochner, \emph{{Determination of the CKM matrix element |Vcb|, the B
  $\to$ X(s) $\gamma$ decay rate, and the b-quark mass}}.
\newblock PhD thesis, Humboldt U., Berlin, 2011.

\bibitem{Tantalo:2013maa}
N.~Tantalo, \emph{{Isospin Breaking Effects on the Lattice}}, {\emph{PoS} {\bf
  LATTICE2013} (2014) 007}, [\href{https://arxiv.org/abs/1311.2797}{{\tt
  1311.2797}}].

\bibitem{Portelli:2015wna}
A.~Portelli, \emph{{Inclusion of isospin breaking effects in lattice
  simulations}}, {\emph{PoS} {\bf LATTICE2014} (2015) 013},
  [\href{https://arxiv.org/abs/1505.07057}{{\tt 1505.07057}}].

\bibitem{Ishikawa:2012ix}
T.~Ishikawa, T.~Blum, M.~Hayakawa, T.~Izubuchi, C.~Jung and R.~Zhou,
  \emph{{Full QED+QCD low-energy constants through reweighting}},
  \href{http://dx.doi.org/10.1103/PhysRevLett.109.072002}{\emph{Phys. Rev.
  Lett.} {\bf 109} (2012) 072002}, [\href{https://arxiv.org/abs/1202.6018}{{\tt
  1202.6018}}].

\bibitem{Aoki:2012st}
S.~Aoki et~al., \emph{{1+1+1 flavor QCD + QED simulation at the physical
  point}}, \href{http://dx.doi.org/10.1103/PhysRevD.86.034507}{\emph{Phys.
  Rev.} {\bf D86} (2012) 034507}, [\href{https://arxiv.org/abs/1205.2961}{{\tt
  1205.2961}}].

\bibitem{deDivitiis:2011eh}
G.~M. de~Divitiis et~al., \emph{{Isospin breaking effects due to the up-down
  mass difference in Lattice QCD}},
  \href{http://dx.doi.org/10.1007/JHEP04(2012)124}{\emph{JHEP} {\bf 04} (2012)
  124}, [\href{https://arxiv.org/abs/1110.6294}{{\tt 1110.6294}}].

\bibitem{deDivitiis:2013xla}
{\scshape RM123} collaboration, G.~M. de~Divitiis, R.~Frezzotti, V.~Lubicz,
  G.~Martinelli, R.~Petronzio, G.~C. Rossi et~al., \emph{{Leading isospin
  breaking effects on the lattice}},
  \href{http://dx.doi.org/10.1103/PhysRevD.87.114505}{\emph{Phys. Rev.} {\bf
  D87} (2013) 114505}, [\href{https://arxiv.org/abs/1303.4896}{{\tt
  1303.4896}}].

\bibitem{Gockeler:1989wj}
M.~Gockeler, R.~Horsley, E.~Laermann, P.~E.~L. Rakow, G.~Schierholz, R.~Sommer
  et~al., \emph{{{QED}: A Lattice Investigation of the Chiral Phase Transition
  and the Nature of the Continuum Limit}},
  \href{http://dx.doi.org/10.1016/0550-3213(90)90490-5}{\emph{Nucl. Phys.} {\bf
  B334} (1990) 527--558}.

\bibitem{Hayakawa:2008an}
M.~Hayakawa and S.~Uno, \emph{{QED in finite volume and finite size scaling
  effect on electromagnetic properties of hadrons}},
  \href{http://dx.doi.org/10.1143/PTP.120.413}{\emph{Prog. Theor. Phys.} {\bf
  120} (2008) 413--441}, [\href{https://arxiv.org/abs/0804.2044}{{\tt
  0804.2044}}].

\bibitem{Endres:2015gda}
M.~G. Endres, A.~Shindler, B.~C. Tiburzi and A.~Walker-Loud, \emph{{Massive
  photons: an infrared regularization scheme for lattice QCD+QED}},
  \href{http://dx.doi.org/10.1103/PhysRevLett.117.072002}{\emph{Phys. Rev.
  Lett.} {\bf 117} (2016) 072002},
  [\href{https://arxiv.org/abs/1507.08916}{{\tt 1507.08916}}].

\bibitem{Polley:1990tf}
L.~Polley and U.~Wiese, \emph{{Monopole condensate and monopole mass in U(1)
  lattice gauge theory}},
  \href{http://dx.doi.org/10.1016/0550-3213(91)90380-G}{\emph{Nucl.Phys.} {\bf
  B356} (1991) 629--654}.

\bibitem{Lucini:2015hfa}
B.~Lucini, A.~Patella, A.~Ramos and N.~Tantalo, \emph{{Charged hadrons in local
  finite-volume QED+QCD with C$^\star$ boundary conditions}},
  \href{http://dx.doi.org/10.1007/JHEP02(2016)076}{\emph{JHEP} {\bf 02} (2016)
  076}, [\href{https://arxiv.org/abs/1509.01636}{{\tt 1509.01636}}].

\bibitem{Fodor:2015pna}
Z.~Fodor, C.~Hoelbling, S.~D. Katz, L.~Lellouch, A.~Portelli, K.~K. Szabo
  et~al., \emph{{Quantum electrodynamics in finite volume and nonrelativistic
  effective field theories}},
  \href{http://dx.doi.org/10.1016/j.physletb.2016.01.047}{\emph{Phys. Lett.}
  {\bf B755} (2016) 245--248}, [\href{https://arxiv.org/abs/1502.06921}{{\tt
  1502.06921}}].

\bibitem{Creutz:1978xw}
M.~Creutz, \emph{{Quantum Electrodynamics in the Temporal Gauge}},
  \href{http://dx.doi.org/10.1016/0003-4916(79)90365-8}{\emph{Annals Phys.}
  {\bf 117} (1979) 471}.

\bibitem{Bloch:1937pw}
F.~Bloch and A.~Nordsieck, \emph{{Note on the Radiation Field of the
  electron}}, \href{http://dx.doi.org/10.1103/PhysRev.52.54}{\emph{Phys. Rev.}
  {\bf 52} (1937) 54--59}.

\bibitem{StefanoI}
S.~Weinberg, \emph{{The Quantum Theory of Field}}, vol.~1. Foundations.
\newblock Cambridge University Press, 2005.

\bibitem{Yennie:1961ad}
D.~R. Yennie, S.~C. Frautschi and H.~Suura, \emph{{The infrared divergence
  phenomena and high-energy processes}},
  \href{http://dx.doi.org/10.1016/0003-4916(61)90151-8}{\emph{Annals Phys.}
  {\bf 13} (1961) 379--452}.

\bibitem{Grammer:1973db}
G.~Grammer, Jr. and D.~R. Yennie, \emph{{Improved treatment for the infrared
  divergence problem in quantum electrodynamics}},
  \href{http://dx.doi.org/10.1103/PhysRevD.8.4332}{\emph{Phys. Rev.} {\bf D8}
  (1973) 4332--4344}.

\bibitem{Weinberg:1965nx}
S.~Weinberg, \emph{{Infrared photons and gravitons}},
  \href{http://dx.doi.org/10.1103/PhysRev.140.B516}{\emph{Phys. Rev.} {\bf 140}
  (1965) B516--B524}.

\bibitem{Lee:1964is}
T.~D. Lee and M.~Nauenberg, \emph{{Degenerate Systems and Mass Singularities}},
  \href{http://dx.doi.org/10.1103/PhysRev.133.B1549}{\emph{Phys. Rev.} {\bf
  133} (1964) B1549--B1562}.

\bibitem{Low:1954kd}
F.~Low, \emph{{Scattering of light of very low frequency by systems of spin
  1/2}}, \href{http://dx.doi.org/10.1103/PhysRev.96.1428}{\emph{Phys.Rev.} {\bf
  96} (1954) 1428--1432}.

\bibitem{GellMann:1954kc}
M.~Gell-Mann and M.~Goldberger, \emph{{Scattering of low-energy photons by
  particles of spin 1/2}},
  \href{http://dx.doi.org/10.1103/PhysRev.96.1433}{\emph{Phys.Rev.} {\bf 96}
  (1954) 1433--1438}.

\bibitem{Becirevic:2009aq}
D.~Becirevic, B.~Haas and E.~Kou, \emph{{Soft Photon Problem in Leptonic
  B-decays}},
  \href{http://dx.doi.org/10.1016/j.physletb.2009.10.017}{\emph{Phys. Lett.}
  {\bf B681} (2009) 257--263}, [\href{https://arxiv.org/abs/0907.1845}{{\tt
  0907.1845}}].

\end{thebibliography}\endgroup

\end{document}